\documentclass[%
 reprint,
amsmath,amssymb,
aps,
pra,
floatfix,
]{revtex4-2}

\usepackage{graphicx}
\usepackage{dcolumn}
\usepackage{bm}
\usepackage{braket}
\usepackage{subcaption}
\usepackage{cleveref}
\usepackage{xcolor}
\begin{document}

\title{Experimental Optimization of a Cryogenic Buffer Gas Source for Molecular Magneto-optical Traps}
\author{Kaiya Wilson}
\author{Ebram Youssef}
\author{Loïc Anderegg} \email{anderegg@usc.edu}

\affiliation{Department of Physics and Astronomy, University of Southern California, Los Angeles, California 90089, USA}%

\date{\today}

\begin{abstract}

Molecular magneto-optical traps (MOTs) are loaded from cryogenic buffer gas beam (CBGB) sources, which produce a large flux of cold molecules. However, only a small fraction of this flux is slowed and captured. Here, we investigate and optimize one of these CBGB sources through a combination of experiment and simulation to maximize the number of molecules loaded into a MOT of CaF molecules. We experimentally determine the range of molecule velocities from the source that are successfully slowed and captured in the MOT, and perform a systematic study of buffer gas source geometries to identify designs that maximize MOT loading. Our observations agree well with Monte Carlo simulations, which provide physical insight into the dynamics within the source and explain how source geometry influences the subset of molecules that are ultimately trapped. These results provide practical design guidelines for optimizing cryogenic buffer gas beam sources used in laser cooling experiments.

\end{abstract}

\maketitle

\section{Introduction}
\label{introduction}
Ultracold molecules have a wide range of applications in precision measurement~\cite{DeMille2024,Hutzler2020}, quantum information and simulation~\cite{DeMille2002,Kaufman2021,Micheli2006,Cornish2024}, and ultracold chemistry~\cite{Bell2009}. In particular, laser cooled molecules offer several important advantages, including nondestructive optical detection and the ability to cool and trap increasingly complex polyatomic species~\cite{Augenbraun2023,Vilas2024,Hallas2026}.

Recent advances in laser cooling and trapping have led to significant improvements in the number of trapped molecules \cite{Fitch2021,Lu2022,Burau2023,Jorapur2024,Li2025,Yu2026,Youssef2026,Nasir2026}. The first step in any such experiment is the production of a cold molecular beam. However, the efficient production of a high-flux cold molecular beam remains a major bottleneck for trapped molecule numbers. Molecules must be produced at sufficiently low temperatures such that most of the population occupies the low lying vibrational and rotational states addressed by the cooling lasers. This becomes even more important with larger polyatomic molecules, due to their additional bending and stretching modes. The workhorse technique for producing such beams is the cryogenic buffer gas beam (CBGB) source~\cite{Hutzler2012}.

Over the past two decades, considerable experimental and theoretical effort has been devoted to understanding and optimizing CBGB sources~\cite{Lu2011, Bulleid2013, Takahashi2021, Skoff2011,  Wright2022, Truppe2017buffer, Singh2018, White2024, Vogeley2025}. Most optimization studies have focused on maximizing beam brightness while minimizing the forward velocity of the extracted beam. Although optimizing these parameters is important for some applications, it does not always directly correlate to the number of molecules that can be captured in a MOT. To date, no studies have systematically investigated how the geometry of a buffer gas beam source affects the MOT loading efficiency. Furthermore, it remains unclear which portion of the initial molecular velocity distribution is ultimately slowed and captured in the MOT.

In this work, we experimentally determine the initial velocity distribution of molecules that are successfully loaded into the MOT. We then perform a systematic experimental study of a selection of buffer gas cell geometries to identify those that maximize the number of trapped molecules. Finally, we use direct simulation Monte Carlo (DSMC) simulations to investigate the dynamics within these sources. The simulations show good agreement with our measurements at higher effective helium flows and provide physical insight into how source geometry influences beam properties and MOT loading. These results establish practical design guidelines for optimizing cryogenic buffer gas beam sources for laser cooling experiments. By increasing the number of trapped molecules, this work could improve the sensitivity of future precision measurements~\cite{Safronova2018,Hutzler2020}, facilitate progress toward laser cooled molecular Bose Einstein condensates~\cite{Bigagli2024,Gadway2016,Langen2025,Schindewolf2025}, and enable larger optical tweezer arrays of molecules~\cite{Anderegg2019,Zhang2022,Holland2023,Vilas2024,Kaufman2021}.

\section{Experimental Setup}
\label{experiment setup}
The experimental apparatus consists of a cryogenic buffer gas beam source, which produces a molecular beam that is laser slowed and trapped in a MOT. The slowing and trapping methods used here were previously described in detail in Ref.~\cite{Youssef2026}. We briefly review these methods and discuss the CBGB source in detail here.

\subsection{Cryogenic Buffer Gas Beam Source}
\label{CBGB source}
\begin{figure}
    \centering
    \includegraphics[width=0.8\linewidth]{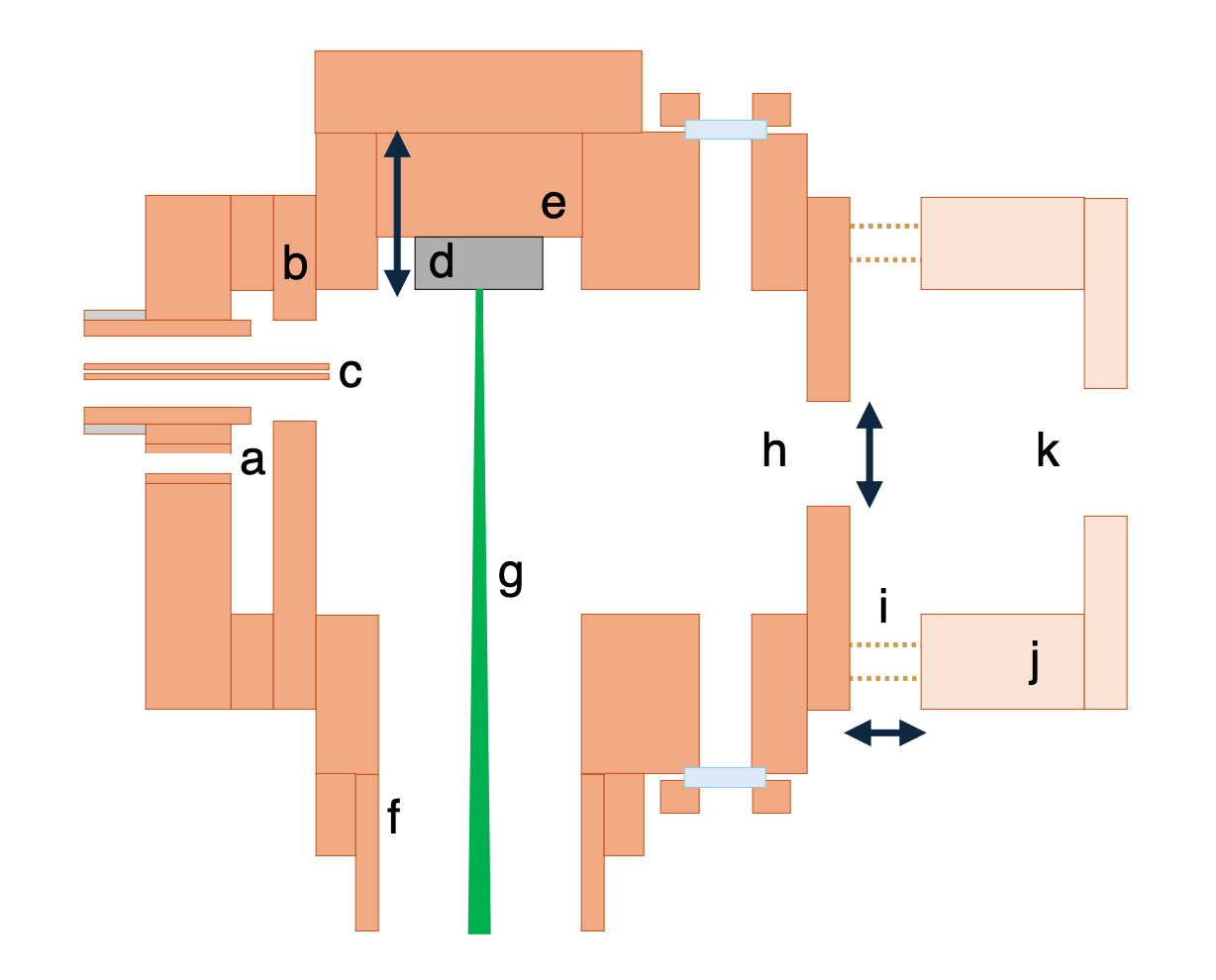}
    \caption{\textbf{Cross-Section of the Cryogenic Cell.} Cell components are as follows: (a) helium fill line, (b) diffuser, (c) $\mathrm{SF}_6$ fill line, (d) calcium target, (e) target holder, (f) snorkel, (g) 532\,\mbox{nm} ablation beam, (h) single-stage cell aperture, (i) brass rods, (j) two-stage cell extender, (k) two-stage cell aperture. Black arrows indicate parameters that are scanned in the experiment. Transparent regions show two-stage geometry. The cell has a 1\,\mbox{in} (2.5\,\mbox{cm}) bore and an internal length of 1.5\,\mbox{in} (3.8\,\mbox{cm}).}
    \label{fig:cell}
\end{figure}

The cryogenic buffer gas beam source consists of a cryogenic copper cell, shown in Fig.~\ref{fig:cell}, in which CaF molecules are produced. The cell is cooled by a pulse tube cryocooler (Cryomech PT310) through a liquid helium damping pot, which reduces temperature oscillations. During operation at a 1\,\mbox{Hz} ablation rate, the cell temperature is 2.4\,\mbox{K}. $\mathrm{SF}_6$ is introduced (at 0.05 - 1 standard cubic centimeters per minute (sccm)) into the cell through a heated fill line at approximately 220\,\mbox{K}. Helium buffer gas (0.5 - 16\,\mbox{sccm}) enters the rear of the cell and passes through a diffuser that distributes the flow uniformly around the circumference of the cell. The geometry of the cell is similar to previous buffer gas sources used for CaF~\cite{Chae2016Thesis,Anderegg2020Thesis,Bao2024Thesis}. The cell has a 1\,\mbox{in} (2.5\,\mbox{cm}) diameter bore and an internal length of 1.5\,\mbox{in} (3.8\,\mbox{cm}). A 1\,\mbox{cm$^2$} square calcium target is mounted opposite a snorkel assembly containing a window through which a 532\,\mbox{nm} pulsed laser is focused onto the target to ablate the calcium. The ablation laser is rastered over the target. It has a pulse duration of 3.9\,\mbox{ns} and the pulse energy can be varied up to 35\,\mbox{mJ}. It is typically operated at $\sim$20\,\mbox{mJ}. 

The source can be operated in either a single-stage or two-stage configuration. In the two-stage configuration, a secondary aperture is positioned downstream of the primary cell exit to provide additional collisions between the molecular beam and the helium buffer gas. This reduces the hydrodynamic boosting and lowers the forward velocity of the extracted beam~\cite{Hutzler2012}. The second stage consists of a 10-mm-diameter aperture mounted at the end of a 13\,\mbox{mm} long cylindrical cell extender. We vary the gap between the single-stage aperture and the second stage.

\subsection{Magneto-optical Trap}
\label{MOT}
Once the molecules exit the cell, they are laser slowed with the $X^2\Sigma^+\rightarrow B^2\Sigma^+$, 531\,\mbox{nm} transition using a beam that is frequency chirped. The molecules are then loaded into a DC MOT operating on the $X^2\Sigma^+\rightarrow A^2\Pi_{1/2}$, 606\,\mbox{nm} transition. The MOT uses a 10:1 ratio of v=1 repump to mainline v=0 power, which was shown to increase the MOT capture velocity~\cite{Youssef2026}. The MOT parameters (intensity, detuning, gradient) were held fixed for all of the results presented in this paper. The delay between the ablation and the start of the slowing as well as the slowing duration were scanned, but the optimal values remained relatively unchanged between cell configurations. The slowing light is applied for 7\,\mbox{ms}, beginning $\sim$2.5\,\mbox{ms} after the ablation pulse. This wait time, as discussed in Sec.~\ref{simulations}, roughly equals the extraction time of the molecules from the cell.

\section{MOT Loading Velocity}
\label{loading velocity}

\begin{figure}[htbp]
    \begin{subfigure}{0.225\textwidth}
        \centering
        \includegraphics[width=\linewidth]{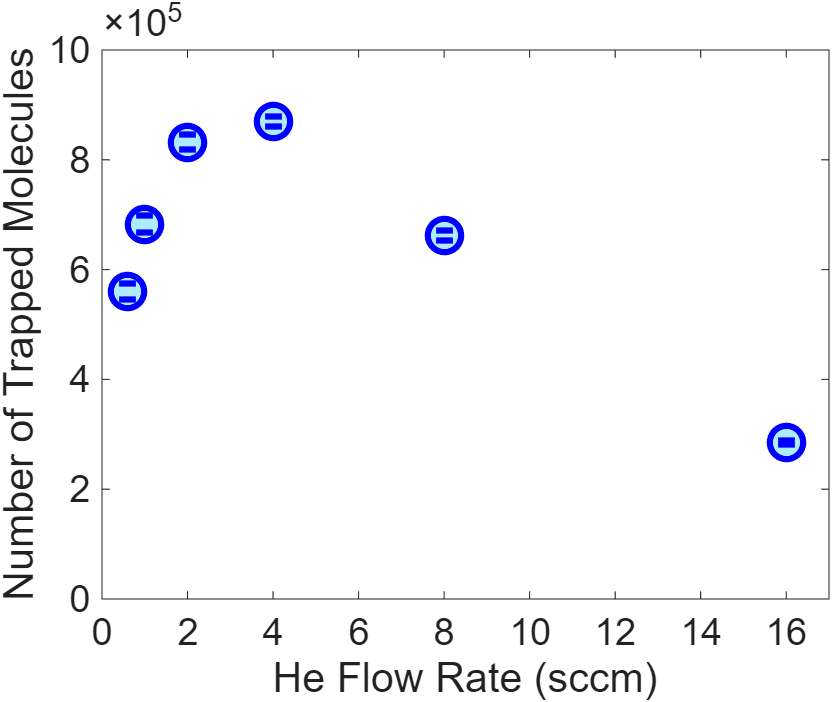}
        \caption{}
        \label{fig:m vs flow example}
    \end{subfigure}
    \begin{subfigure}{0.25\textwidth}
        \centering
        \includegraphics[width=\linewidth]{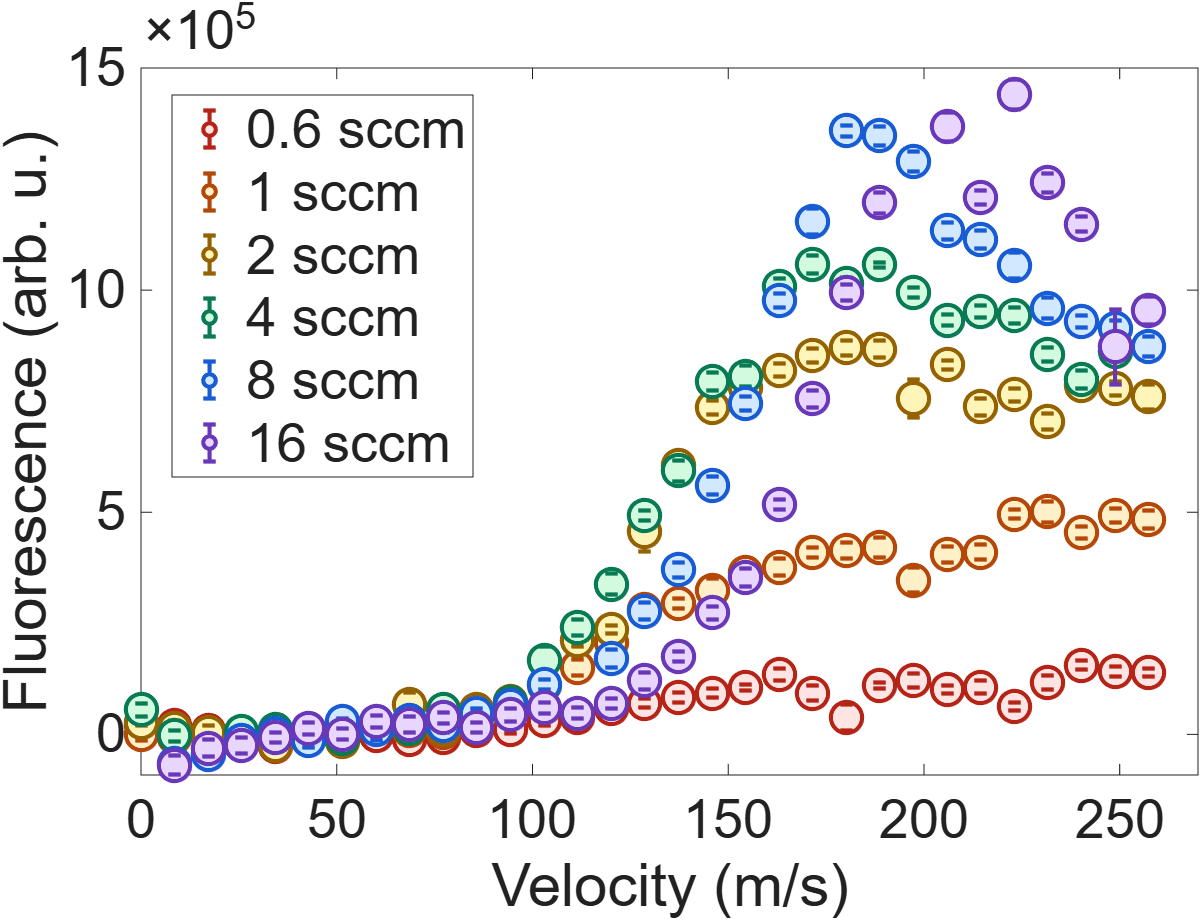}
                \caption{}
        \label{fig:v_scan_example}
    \end{subfigure}
            \begin{subfigure}{0.25\textwidth}
        \centering
        \includegraphics[width=\linewidth]{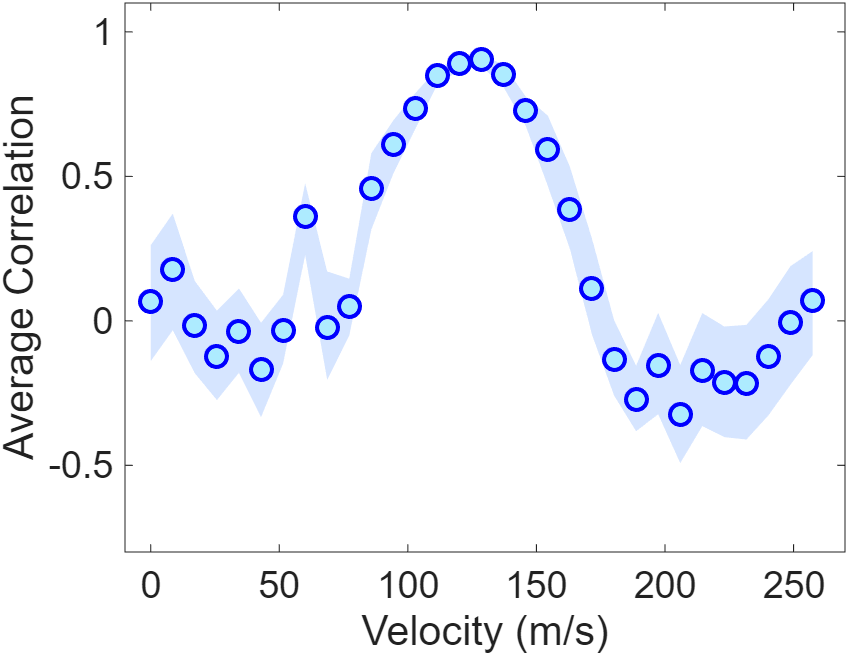}
        \caption{}
        \label{fig:v vs correlation}
    \end{subfigure}
    \caption{\textbf{Initial Loading Velocity.} (a) The number of molecules trapped in the MOT as a function of the continuous helium flow rate for a particular cell configuration. Error bars represent $\pm$1 standard error of the mean statistical uncertainties. (b) Molecular beam fluorescence as a function of beam forward velocity, measured using a Doppler-sensitive laser beam. This data is the corresponding molecular flux scan for the MOT trapping number scan shown in (a). Error bars represent $\pm$1 standard error of the mean statistical uncertainties. (c) Average correlation between the MOT number and the forward velocity scans over all cell configurations. The shaded region indicates 1$\sigma$ statistical uncertainties. When fitted to a Gaussian, the center peak is located at 123\,\mbox{m/s} with a width of 29\,\mbox{m/s}.}
    \label{fig:capture velocity scan}
\end{figure}

In order to compare our simulations with the observed MOT loading trends, we must determine which initial velocities of the molecular beam are loaded into our MOT. To measure this, we employ a two-step process. First, we measure the number of molecules loaded in the MOT as a function of the continuous helium flow rate into the cell, typically 0.6, 1, 2, 4, 8, and 16\,\mbox{sccm} (Fig.~\ref{fig:m vs flow example}). Second, we take a Doppler sensitive forward velocity measurement of the molecular beam as a function of the continuous helium flow rate (Fig.~\ref{fig:v_scan_example}). The latter is collected by scanning the frequency of the X-A transition laser with a 6\,\mbox{mm} diameter detection beam at 45 degrees with respect to the molecular beam. This provides a measure for the number of molecules produced by the source for each velocity.

For each velocity class, we then compute the correlation between the beam signal and the MOT number across the set of helium flows. We repeat this process for numerous cell geometries, many of which are discussed in the following sections. In Fig.~\ref{fig:v vs correlation}, we show the correlation between these two scans as a function of the velocity of the molecules. We define the correlation by the Pearson correlation coefficient (Appendix \ref{supp:correlation}). The correlation peaks at $\sim$123\,\mbox{m/s} with a maximum correlation of 0.95. A Gaussian fit gives a width of $\sigma=29$\,\mbox{m/s}. We therefore take a window of this width centered on the peak, 108.5 - 137.5\,\mbox{m/s}, as the range of initial velocities before slowing that are loaded into the MOT for all simulations.

These results can provide additional insight into our slowing process; given an initial molecular velocity centered at 123\,\mbox{m/s} and a slowing duration of 7\,\mbox{ms}, the average scattering rate during slowing is 1.3\,\mbox{MHz}, and the average slowing distance is 0.45\,\mbox{m}. We note that this distance is significantly shorter than the actual distance from the second stage of the cell to the center of the MOT (0.82\,\mbox{m}). This disparity can be partially explained by the spread in pumpout time shown in Fig.~\ref{fig:pumpout}.

We note that the MOT loads molecules from the slow tail of the velocity distribution rather than the peak. However, assuming the derived average scattering rate, slowing the peak velocity of $\sim$180\,\mbox{m/s} would require extending the slowing distance by a factor of $(180/123)^2\approx2$. This would result in a $4\times$ decrease in the fraction of molecules that reach the MOT due to beam divergence, as well as increased transverse pluming from slowing. We note that these factors may change if transverse cooling~\cite{Nasir2026} were implemented.

\section{Simulations}
\label{simulations}
To simulate the dynamics of the cryogenic buffer gas source, we follow a two step approach, as done in~\cite{Takahashi2021}. First, the steady-state flow of the helium buffer gas is simulated using DSMC. Then, the trajectories of CaF molecules are calculated as they collide with the helium. The DSMC simulation, run with DS2V~\cite{Bird2013}, is assumed to be axially symmetric. The simulation geometry matches the experimental apparatus, with the temperature of the cell walls set to 2.5\,\mbox{K}. The helium-helium collisions in the DSMC are modeled as a variable hard sphere, with a reference diameter of $2.3\times 10^{-10}$m at 273\,\mbox{K} and viscosity $\omega=0.66$. The CaF-helium collisions are treated as hard-sphere collisions with a cross-section of $1.0\times 10^{-14}\mathrm{cm}^2$~\cite{Maussang2005}. We use a dynamic timestep of 0.2/$\Gamma$~\cite{Takahashi2021}, where $\Gamma = \sigma n_{He}\langle v_{rel}\rangle$. Here, $\sigma$ is the CaF-helium cross-section, $n_{He}$ is the number density of the buffer gas and $v_{rel}$ is the relative velocity between the CaF molecule and a randomly chosen helium atom.

\begin{figure}[htbp]
    \centering

    \begin{subfigure}[t]{0.48\columnwidth}
        \centering
        \includegraphics[width=\linewidth]{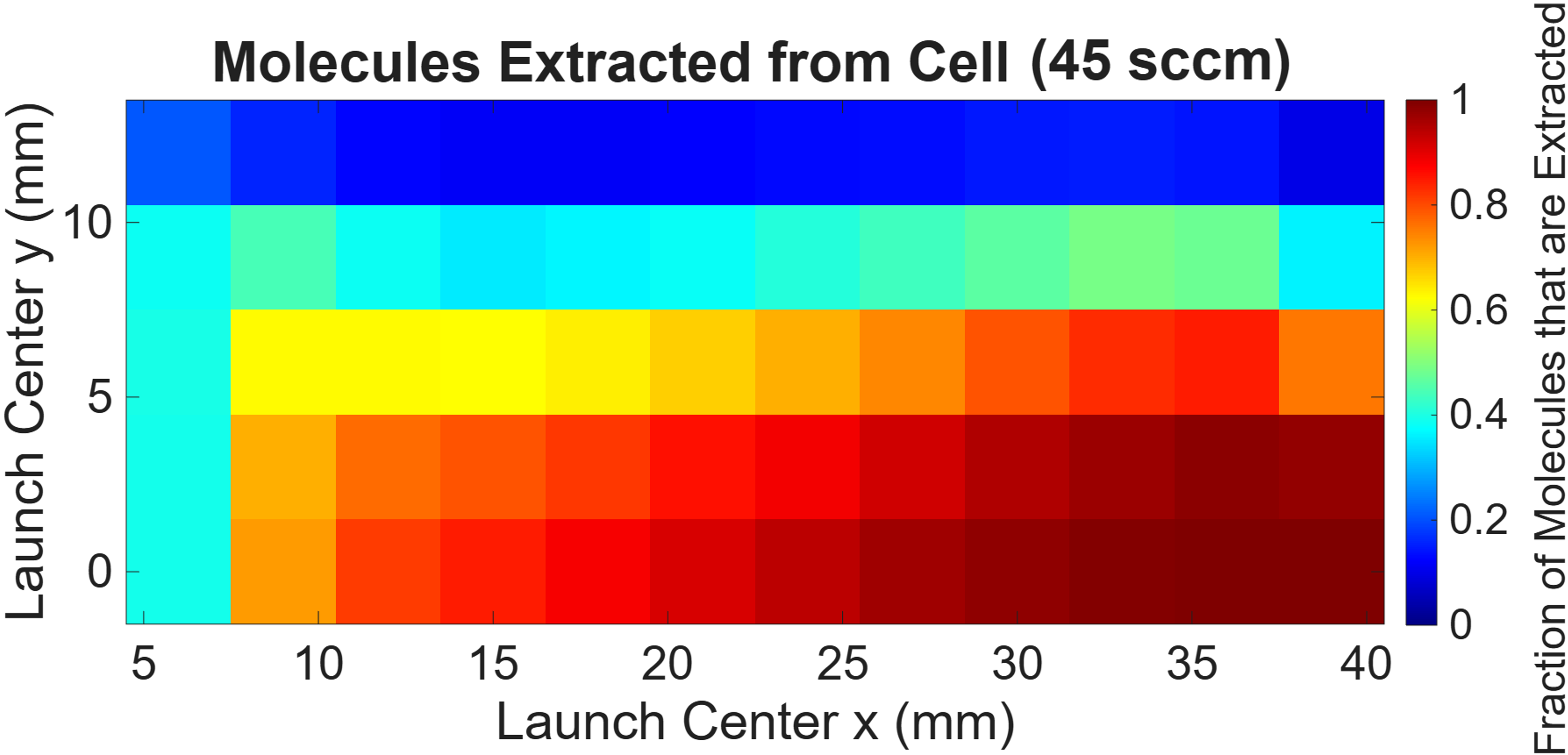}
        \caption{}
        \label{fig:extractionlow}
    \end{subfigure}
    \hfill
    \begin{subfigure}[t]{0.50\columnwidth}
        \centering
        \includegraphics[width=\linewidth]{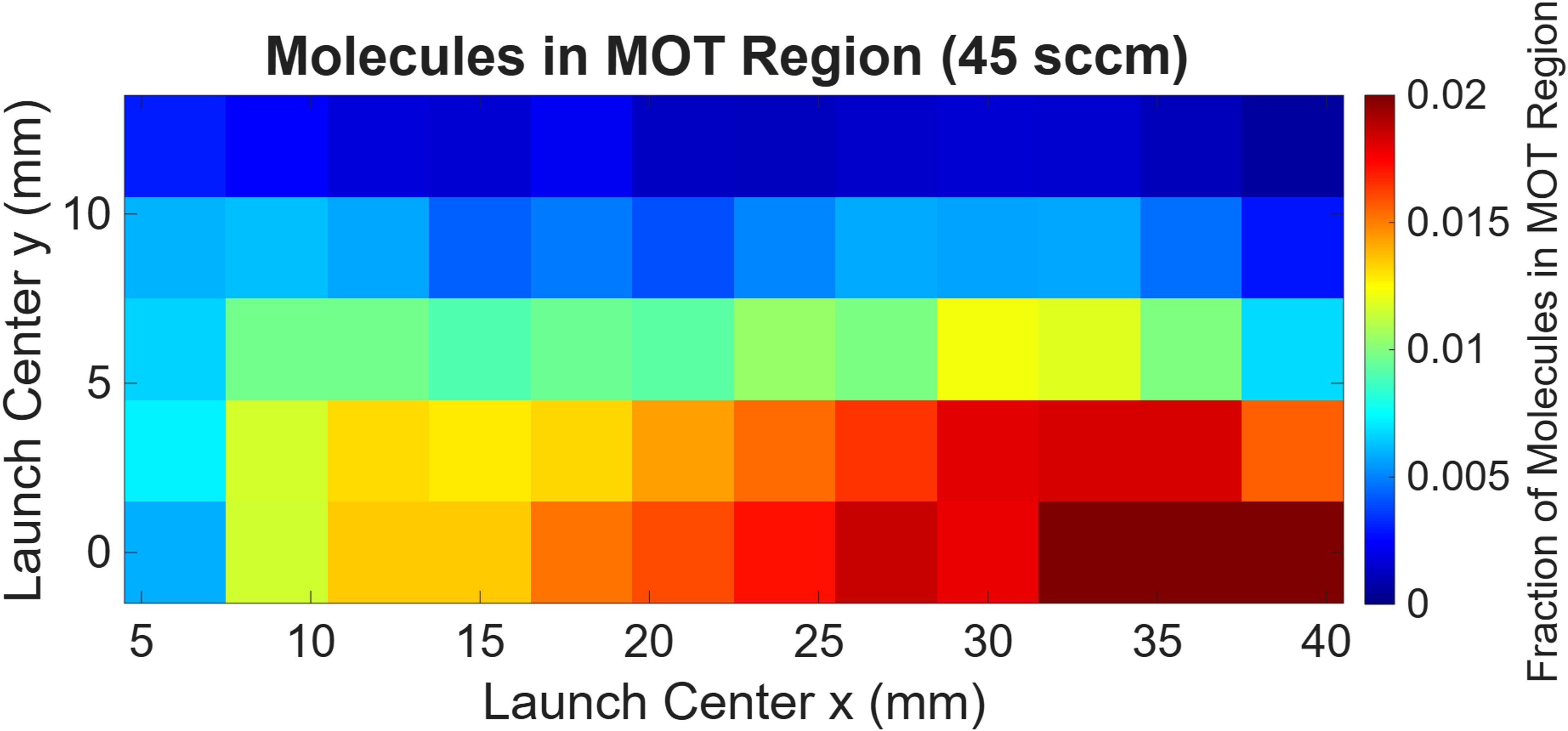}
        \caption{}
        \label{fig:detectionlow}
    \end{subfigure}
   
    \medskip

\begin{subfigure}[t]{0.48\columnwidth}
        \centering
        \includegraphics[width=\linewidth]{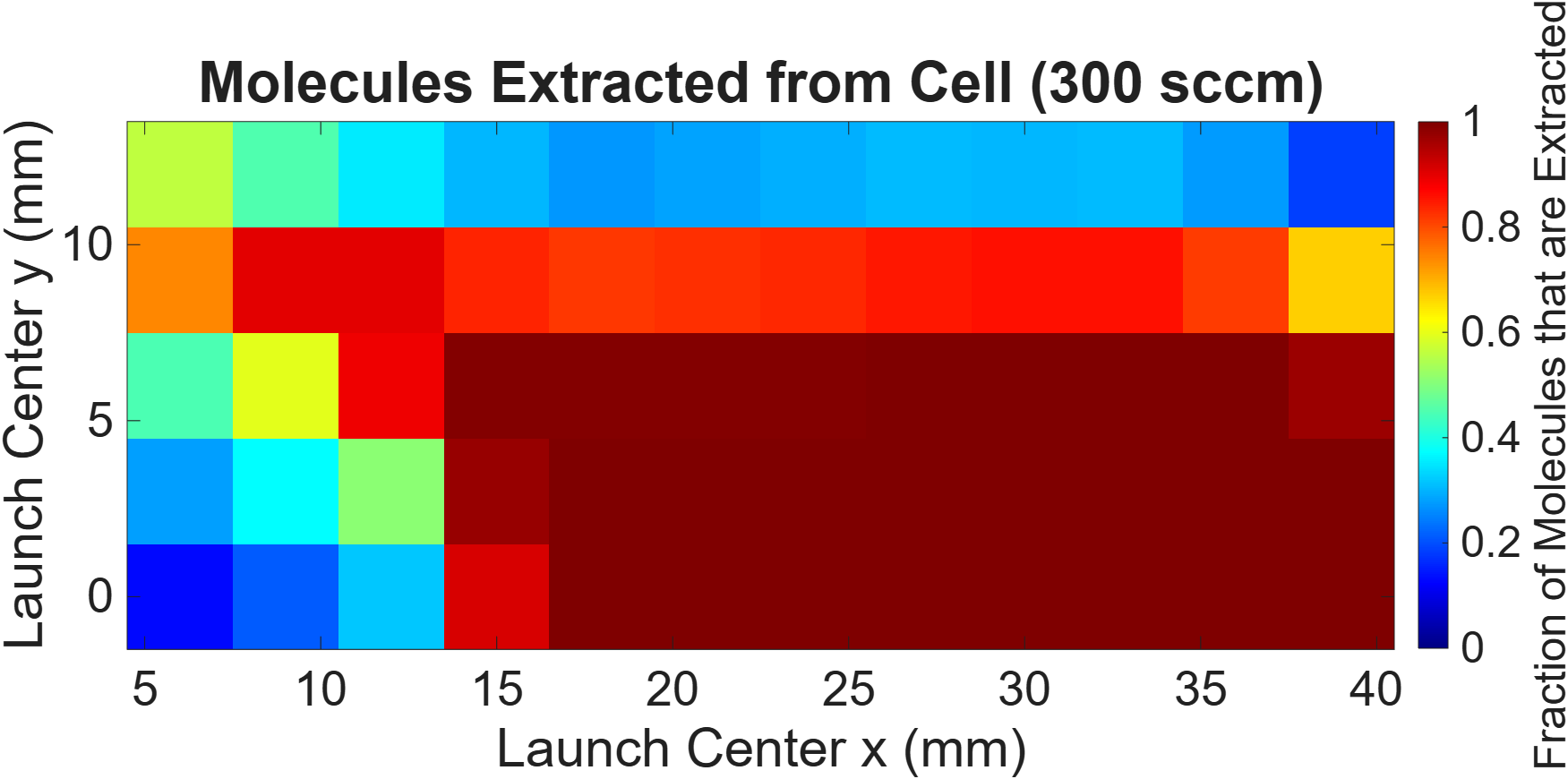}
        \caption{}
        \label{fig:extractionhigh}
    \end{subfigure}
    \hfill
    \begin{subfigure}[t]{0.50\columnwidth}
        \centering
        \includegraphics[width=\linewidth]{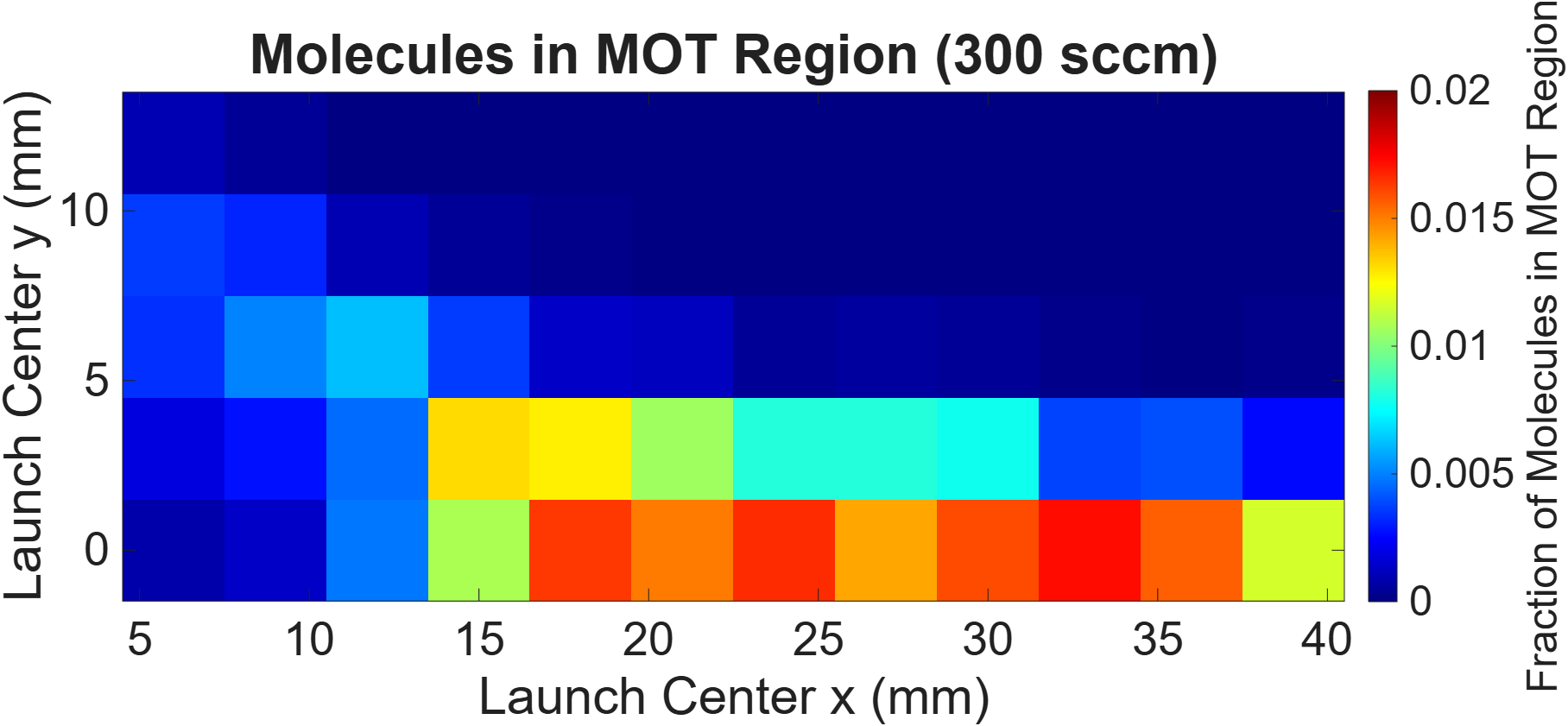}
        \caption{}
        \label{fig:detectionhigh}
    \end{subfigure}

    \medskip

    \begin{subfigure}{0.80\columnwidth}
        \centering
        \includegraphics[width=\linewidth]{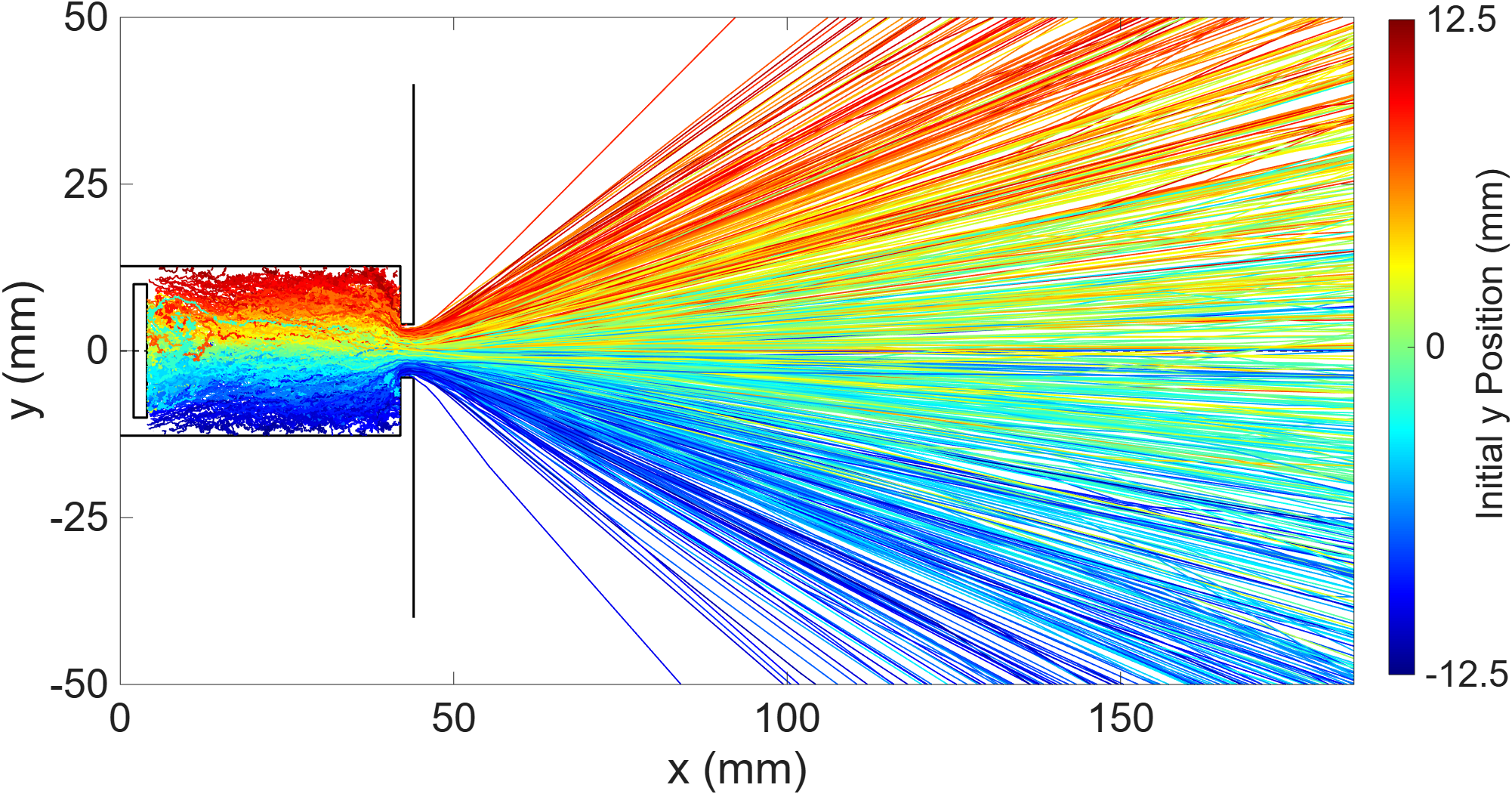}
        \caption{}
        \label{fig:colorplot}
    \end{subfigure}

\caption{\textbf{CaF Extraction.} (a) Fraction of launched CaF molecules that exit the single-stage cell aperture as a function of their initial positions. Here, we use a single-stage cell geometry with an 8\,\mbox{mm} aperture and a 45\,\mbox{sccm} effective (equivalent to 1.5\,\mbox{sccm} continuous) helium flow rate. Helium is flowed in from the back of the cell. (b) Fraction of launched CaF molecules that reach the MOT region as a function of their initial positions. Here, we use a single-stage cell geometry with an 8\,\mbox{mm} aperture and a 45\,\mbox{sccm} effective (equivalent to 1.5\,\mbox{sccm} continuous) helium flow rate. Helium is flowed in from the back of the cell.  (c), (d) Same as (a) and (b) but using a 300\,\mbox{sccm} effective (equivalent to 10\,\mbox{sccm} continuous) helium flow rate. (e) CaF molecule tracks for a 300\,\mbox{scmm} effective helium flow rate, seeded uniformly across the buffer gas cell and colored by initial lateral position.}
    \label{fig:caf-extraction}
\end{figure}

Within the simulation, the CaF particles can be injected into the cell at any given location. Previous simulations and experimental optimizations have investigated the extraction efficiency of buffer gas cells~\cite{Lu2011, Bulleid2013, Takahashi2021, Skoff2011,  Wright2022, Truppe2017buffer, Singh2018, White2024, Vogeley2025}. Here, the extraction efficiency is the fraction of CaF molecules within the cell that exit the single-stage aperture. As we scan the launch location of CaF molecules, we find a relatively uniform extraction, falling away at the edges and back of the cell (Figures~\ref{fig:extractionlow} and~\ref{fig:extractionhigh}). However, if we only consider the molecules that reach the downstream MOT capture region, simulated by adding a 2.5\,\mbox{cm} radius circular detector, 50\,\mbox{cm} downstream from the source (Appendix ~\ref{supp:simulation}),  the spatial dependence changes dramatically  (Fig.~\ref{fig:detectionlow}). Only molecules produced near the cell axis reach the MOT. This is due to the laminar nature of the buffer gas flows in this flow regime.

At higher flows, this effect becomes more pronounced (Fig.~\ref{fig:detectionhigh}). This indicates that it may be experimentally valuable to attempt molecule production in this region of the cell. We note that we assume ballistic trajectories from the source to the MOT, and do not take into account the effects of laser slowing or the resulting transverse heating. While we are unable to experimentally scan the production location directly, we observe an effect from moving the position of the calcium target, as discussed in Sec.~\ref{calcium target position}.

Additionally, we can examine this behavior by simulating particles launched throughout the entire cell and plotting their trajectories (Fig.~\ref{fig:colorplot}). The molecules that are created in the top half of the cell expand upwards while those in the lower half travel downwards, with only a narrow production area in which the molecules have a chance to reach the MOT region.

We benchmark these simulations by comparing their results to the experiment. First, we compare the number of molecules that are trapped in the MOT versus the helium flow rate. Simulations produce molecules that reach the MOT region with many velocities, but we select the velocity class that we determined are loaded in the MOT (108.5 - 137.5\,\mbox{m/s}) and only display those molecules (Appendix~\ref{supp:simulation}). 

\begin{figure}[htbp]
    \centering
    \begin{subfigure}{0.46\columnwidth}
        \centering       \includegraphics[width=\linewidth]{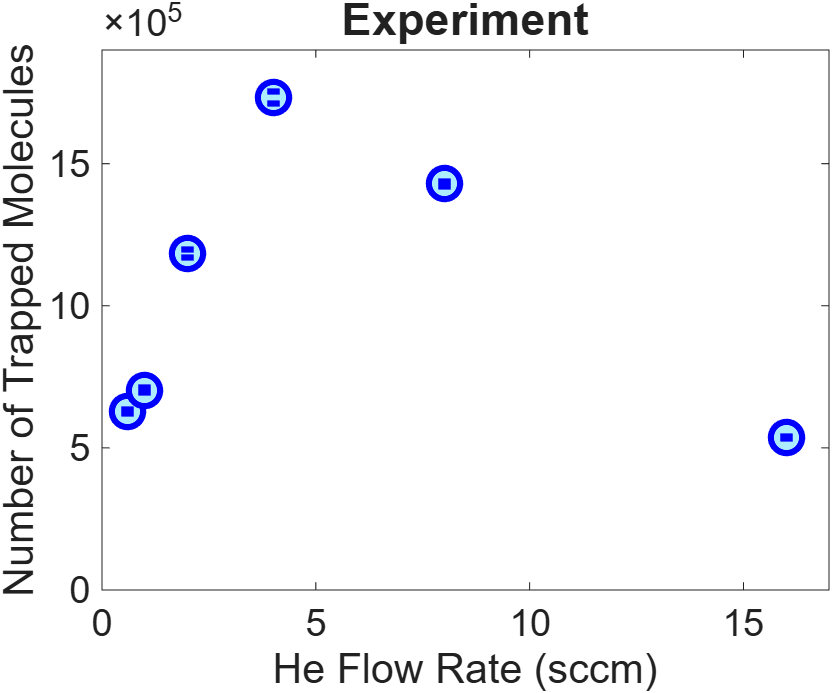}
        \caption{}
        \label{fig:mot scan experiment}
    \end{subfigure}
    \begin{subfigure}{0.485\columnwidth}
        \centering        \includegraphics[width=\linewidth]{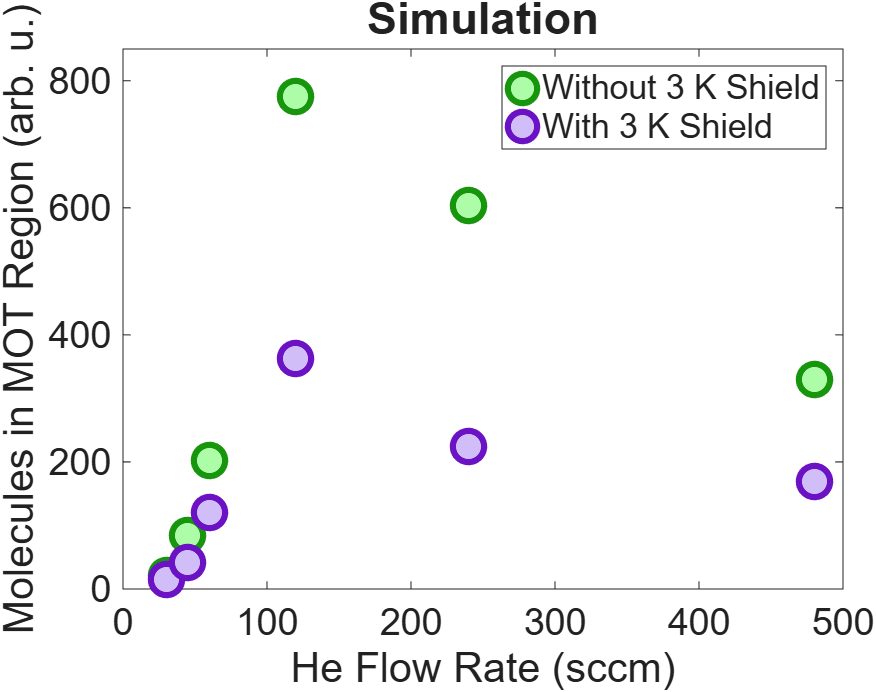}
        \caption{}
        \label{fig:mot scan simulation}
    \end{subfigure}
    \begin{subfigure}{0.7\columnwidth}
        \centering
        \includegraphics[width=\linewidth]{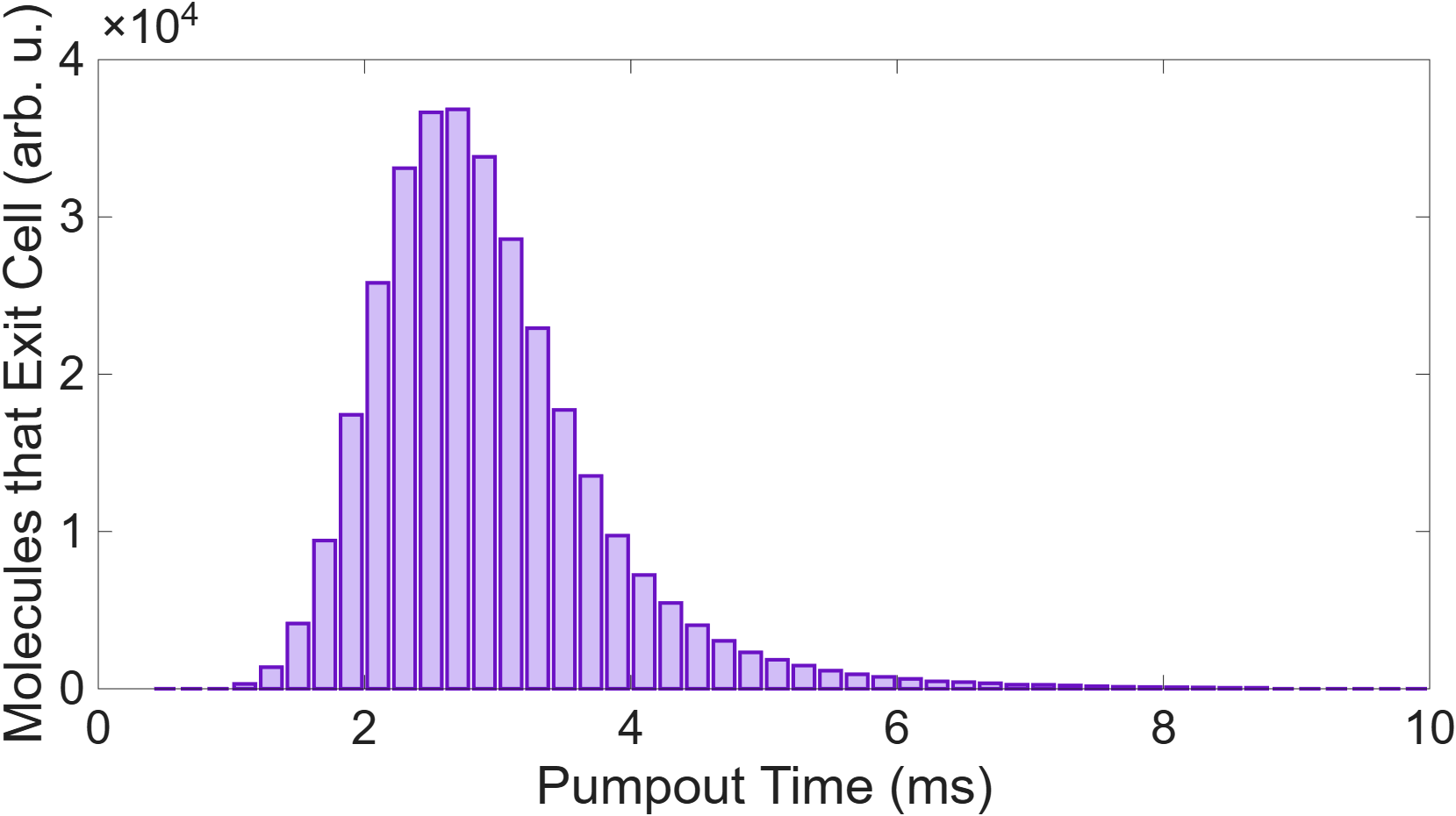}
        \caption{}
        \label{fig:pumpout}
    \end{subfigure}
    
    \caption{\textbf{Simulations Benchmark.} (a) Experimental trapped molecule number versus continuous helium flow rate. This scan is performed with an 8\,\mbox{mm} single-stage cell aperture and a 5.5\,\mbox{mm} two-stage gap. Error bars represent $\pm$1 standard error of the mean statistical uncertainties. (b) Simulated number of molecules in the 108.5 - 137.5\,\mbox{m/s} velocity class that reach the MOT region versus effective helium flow rate. The simulation uses an 8\,\mbox{mm} single-stage cell aperture and a 5.5\,\mbox{mm} two-stage gap. Purple indicates simulation results for a geometry that includes the 13\,\mbox{mm} aperture of the 3\,\mbox{K} shield, while green shows simulation results for a geometry without the 3\,\mbox{K} shield. (c) Simulation results for the distribution of the time it takes for a CaF molecule to exit the cell. CaF molecules are seeded on axis with a Gaussian width of 3\,\mbox{mm}, $\sim$20\,\mbox{mm} from the single-stage aperture, similar to where the calcium target is located experimentally. Results are shown for a two-stage geometry with an 8\,\mbox{mm} single-stage cell aperture and 5.5\,\mbox{mm} two-stage gap.}
    \label{fig:mot scan comparison}
\end{figure}

Fig.~\ref{fig:mot scan comparison} shows experimental and simulation results for the same cell geometry. We find good qualitative agreement with the experimental data when we set the helium flow rate to $\sim$30 times the continuous experiment flow rate (Figures~\ref{fig:mot scan experiment} and~\ref{fig:mot scan simulation}). This increase in effective flow was seen across all configurations of the experiment. We experimentally observe a large increase in pressure in our UHV chamber when we ablate the calcium target, consistent with a higher effective helium flow rate inside the cell. We attribute this to helium that has adsorbed onto the $\mathrm{SF}_6$ ice and ablation dust that builds up inside the cell. The ablation pulse heats these surfaces and releases the adsorbed helium. We note that in our simulations, we model the helium as entering the cell from the back wall behind the diffuser. Based on the aforementioned model, the helium may be released uniformly from the cell walls, however, we do not see a significant change in the simulated extraction results between these two flow settings (Appendix~\ref{supp:simulation}).   

Experimentally, the cell is surrounded by a rectangular enclosure held at 3\,\mbox{K}. This enclosure has a 13\,\mbox{mm} aperture, through which the molecular beam exits, and is located $\sim$15\,\mbox{cm} in front of the single-stage cell aperture. Within the simulation, we can include the effect of collisions due to this aperture, and we find that at higher flows, the presence of the 3\,\mbox{K} shield has a non-negligible impact on the molecular beam measured (Fig.~\ref{fig:mot scan simulation}). The effect of this shield is to reflect some of the helium gas back into the molecular beam. This reflected helium collides with the CaF molecules, knocking them out of the beam that reaches the MOT downstream. At higher flows, due to the smaller mean free path of the helium, this effect becomes larger. In the simulations, the 3\,\mbox{K} shield is set to absorb 25\% of the incoming helium to mimic the layer of coconut charcoal that acts as a cryo-adsorber on this surface. Scanning the absorption from $0\%$ to $50\%$ did not appreciably change the general shape of the resulting trend. As a result, all the simulations presented within this work include the 3\,\mbox{K} shield with a $25\%$ absorption of the helium gas.

The simulation also accurately reproduces the pumpout time of the cell. When launching molecules in the region where the ablation plume is located, we see that the pumpout time peaks around 3\,\mbox{ms} (Fig.~\ref{fig:pumpout}), which agrees with the typical wait time between the ablation and slowing pulses that results in the best MOT number experimentally, 2.5\,\mbox{ms}.

\section{Source Optimization}
\label{source optimization}
Experimental optimization of buffer gas sources is notoriously difficult due to the large number of parameters affecting the molecular beam properties. In this section, we experimentally scan various parameters of the CBGB cell independently to gain an understanding of the general trends present. For each cell configuration, we optimized the flow rates, ablation parameters, and slowing parameters, then record the MOT brightness over the course of $\sim$10 hours, during which the MOT images were taken at a rate of 1\,\mbox{Hz}. We then average over the 80 - 98th percentile of brightness data from that period. This average is used as the comparison metric across the different cell configurations (Appendix~\ref{supp:data selection}). Using this high percentile range allows us to exclude the low MOT signal resulting from changes such as drifts in laser frequencies when imaging unattended, and parameter scans conducted while imaging. This also allows us to exclude data collected during the first $\sim$2 - 5 hours of post-cooldown running, during which we observe a characteristically different molecular beam from the steady state. We find that after this initial period, the source is stable for many days of running.

\subsection{Single-Stage Cell Aperture Size}
\label{cell-aperture size}
\begin{figure}[htbp]
    \begin{subfigure}{0.235\textwidth}
        \centering
        \includegraphics[width=\linewidth]{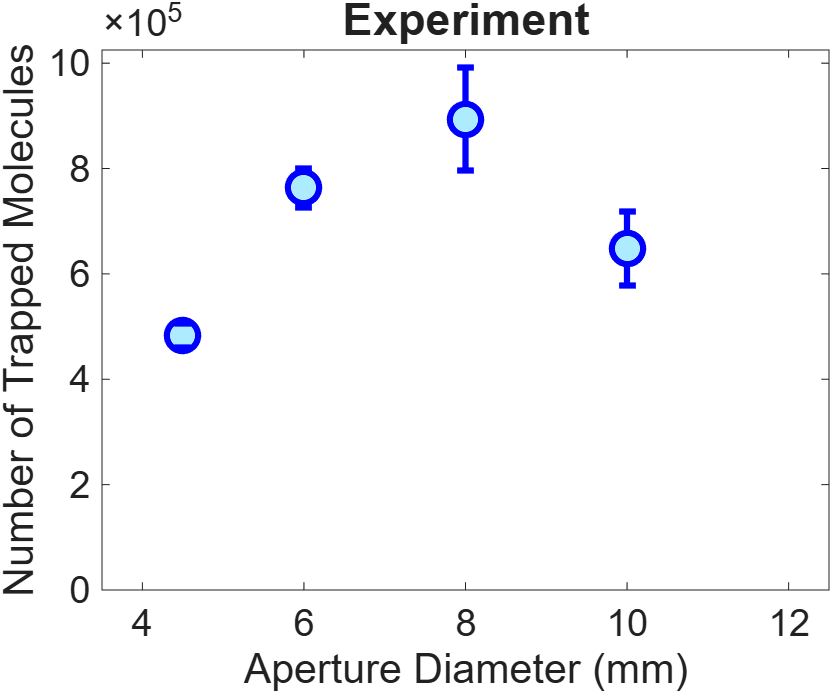}
        \caption{}
        \label{fig:1 stage aperture experiment}
    \end{subfigure}
    \begin{subfigure}{0.24\textwidth}
        \centering
        \includegraphics[width=\linewidth]{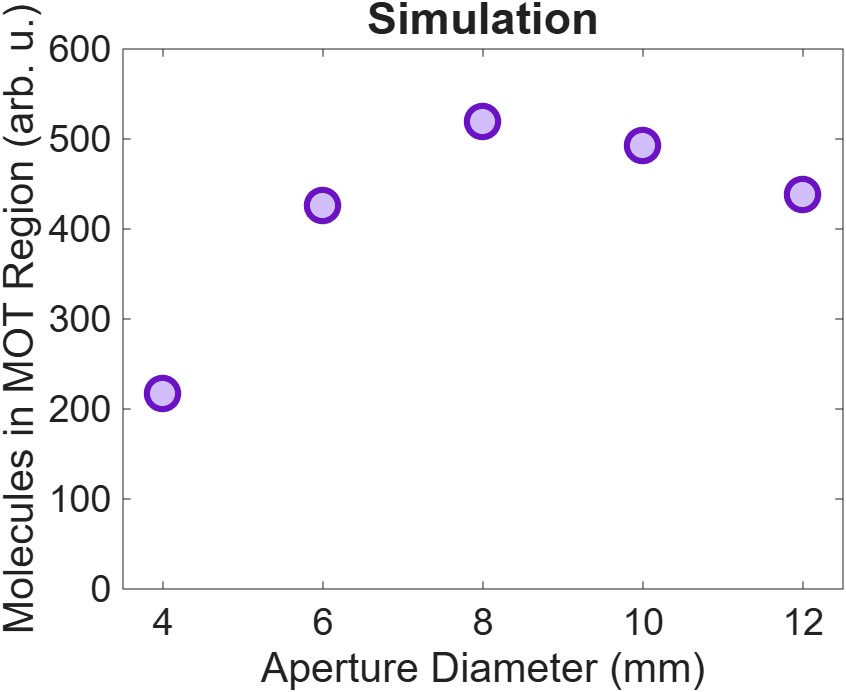}
        \caption{}
        \label{fig:1 stage aperture simulation}
    \end{subfigure}
    \caption{\textbf{Varying Cell Aperture Size.} (a) Average number of trapped molecules for each of four single-stage cell aperture diameters (4.5, 6, 8, and 10\,\mbox{mm}). Trapped molecule number is calculated by averaging the 80 - 98th percentile of each $\sim$10 hour period of imaging for each aperture size. Error bars show the standard deviation of the signal within the selected range. (b) DSMC simulation of 108.5 - 137.5\,\mbox{m/s} velocity molecules that enter the MOT region with an effective helium flow rate of 120\,\mbox{sccm} for 4, 6, 8, 10, and 12\,\mbox{mm} cell aperture diameters.}
    \label{fig:1 stage aperture}
\end{figure}

We began our experimental investigation by scanning the exit aperture diameter of the cell with a single stage. The cell apertures are 1.6\,\mbox{mm} thick, and the diameters tested were 4.5, 6, 8, and 10\,\mbox{mm}.
Results are presented in Fig.~\ref{fig:1 stage aperture experiment}. The MOT brightness peaks for the 8-mm-diameter aperture with an average of 0.9 million molecules trapped in the MOT. This is consistent with the DSMC simulations, which show a peak of molecules with velocities between 108.5 - 137.5\,\mbox{m/s} that reach the MOT region at 8\,\mbox{mm} (Fig.~\ref{fig:1 stage aperture simulation}).

\subsection{Two-Stage Cell Spacing}
\label{2-stage cell}
\begin{figure}[htbp]
    \begin{subfigure}{0.230\textwidth}
        \centering
        \includegraphics[width=\linewidth]{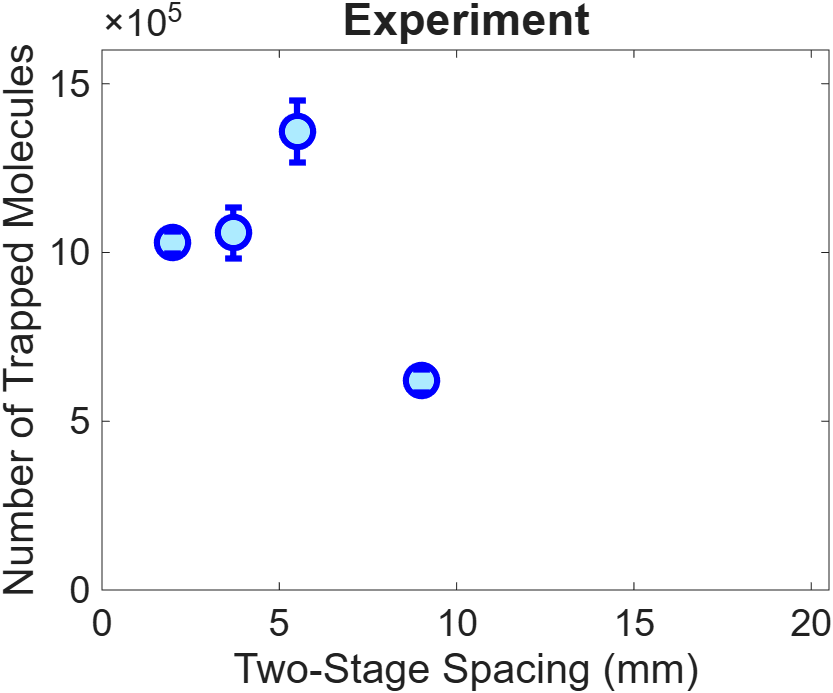}
        \caption{}
        \label{fig:2 stage spacing experiment}
    \end{subfigure}
    \begin{subfigure}{0.235\textwidth}
        \centering        \includegraphics[width=\linewidth]{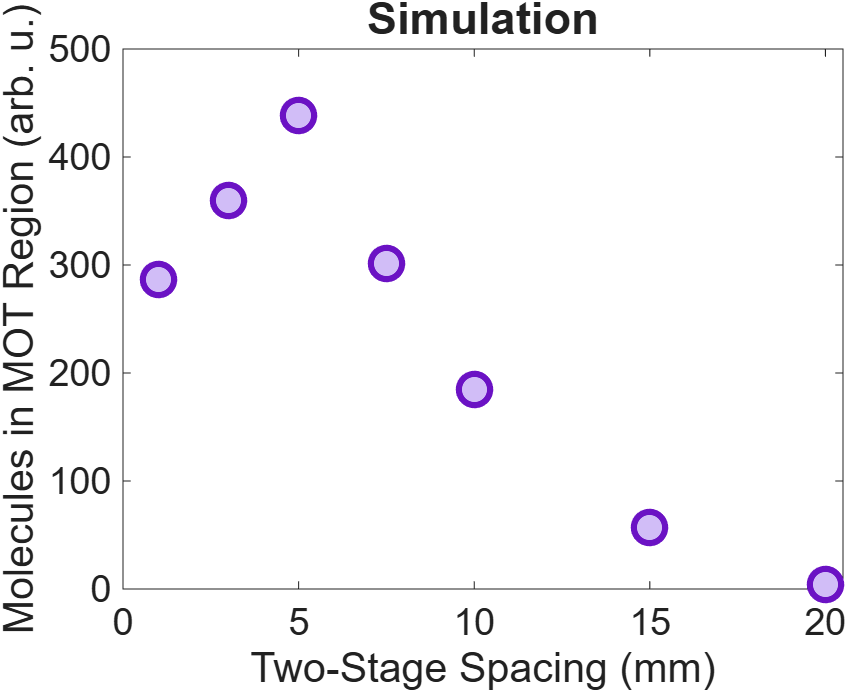}
        \caption{}
        \label{fig:2 stage spacing simulation}
    \end{subfigure}
    \caption{\textbf{Varying Two-Stage Spacing.} (a) Average number of trapped molecules for each of four two-stage cell spacings (2, 3.7, 5.5, and 9\,\mbox{mm}). Trapped molecule number is calculated by averaging the 80 - 98th percentile of each $\sim$10 hour period of imaging for each spacing. Error bars show the standard deviation of the signal within the selected range. (b) DSMC simulation of 108.5 - 137.5\,\mbox{m/s} velocity molecules that enter the MOT region with an effective helium flow rate of 120\,\mbox{sccm} for 1, 3, 5, 7.5, 10, 15, and 20\,\mbox{mm} two-stage cell spacings.}
    \label{fig:2 stage spacing}
\end{figure}

Next, we add a second stage to the cell. The second stage is composed of a copper cell extender with an inner diameter of 1\,\mbox{in} (2.5\,\mbox{cm}) and a length of 13\,\mbox{mm}. The single-stage aperture diameter is set to 8\,\mbox{mm} and the second-stage aperture diameter is set to 10\,\mbox{mm}. We then vary the distance between the single-stage aperture and the extender using copper spacers on four brass rods at the corners of the aperture plate. We chose 2, 3.7, 5.5, and 9\,\mbox{mm} as our spacer lengths. Fig.~\ref{fig:2 stage spacing experiment} shows that, compared with all other two-stage spacings, the 5.5\,\mbox{mm} spacing gave the brightest MOT on average with 1.4 million molecules, $\sim$1.5 times larger than the optimal single-stage source. This agrees very well with simulations (Fig.~\ref{fig:2 stage spacing simulation}), which indicate that a 5\,\mbox{mm} spacing should result in the greatest number of trapped molecules.
 
\subsection{Calcium Target Position}
\label{calcium target position}
\begin{figure}[htbp]
    \centering
    \includegraphics[width=0.5\linewidth]{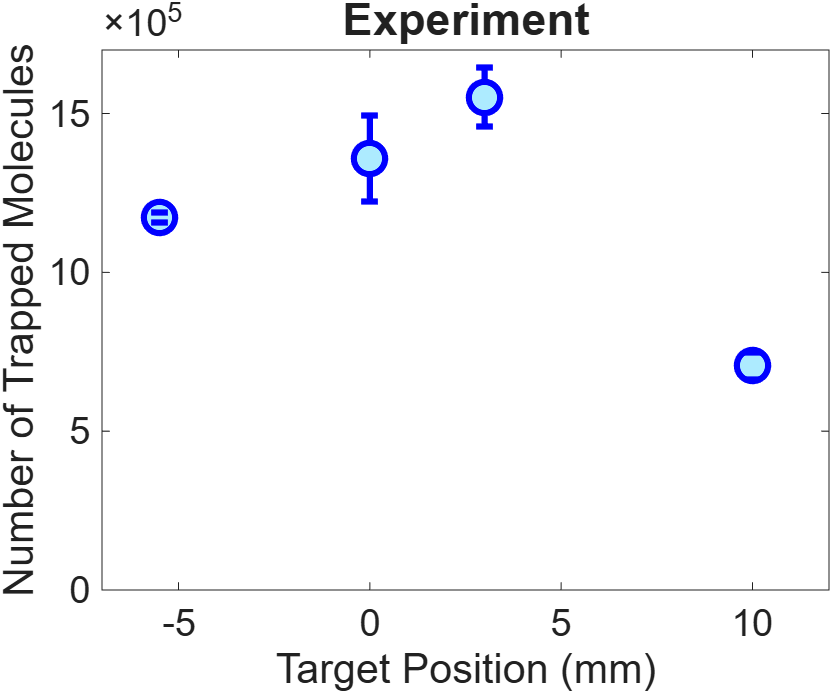}
     \caption{\textbf{Varying Target Position.} Average number of trapped molecules for each of four target holder lengths (-5.5, 0, 3, and 10\,\mbox{mm}), represented by Fig.~\ref{fig:cell}e. Trapped molecule number is calculated by averaging the 80 - 98th percentile of each $\sim$10 hour period of imaging for each target holder length. Error bars show the standard deviation of the signal within the selected range.}
    \label{fig:target position experiment}
\end{figure}

As discussed in Sec.~\ref{simulations}, the simulations (Fig.~\ref{fig:detectionhigh}) suggest that the number of molecules extracted in the selected velocity class depends on the location where the molecules are produced. When the calcium is ablated within the cell, it forms an ablation plume, which reacts with the $\mathrm{SF}_6$ gas to produce the CaF molecules. The exact size of the ablation plume depends on the parameters of the ablation laser, target properties, and the local helium density, as shown in previous studies~\cite{Skoff2011}. While our simulations do not take into account the dynamics related to the ablation process or the three dimensional nature of the helium and $\mathrm{SF}_6$ flows around the square target, we can, nevertheless, study the effect of the target location experimentally. This is done by changing the length of the copper cylinder to which the calcium target is attached, which is indicated by Fig.~\ref{fig:cell}e. A target holder length of ``zero" places the target surface flush with the cell bore, and positive values protrude into the bore. We test target holder lengths of -5.5, 0, 3, and 10\,\mbox{mm}, with the two-stage spacing fixed at 5.5\,\mbox{mm}. Fig.~\ref{fig:target position experiment} shows that the 3\,\mbox{mm} target holder gave the brightest MOT with $\sim$1.6 million molecules.

\subsection{Cell Temperature}
\label{cell temperature}
\begin{figure}[htbp]
    \centering
    \includegraphics[width=0.5\linewidth]{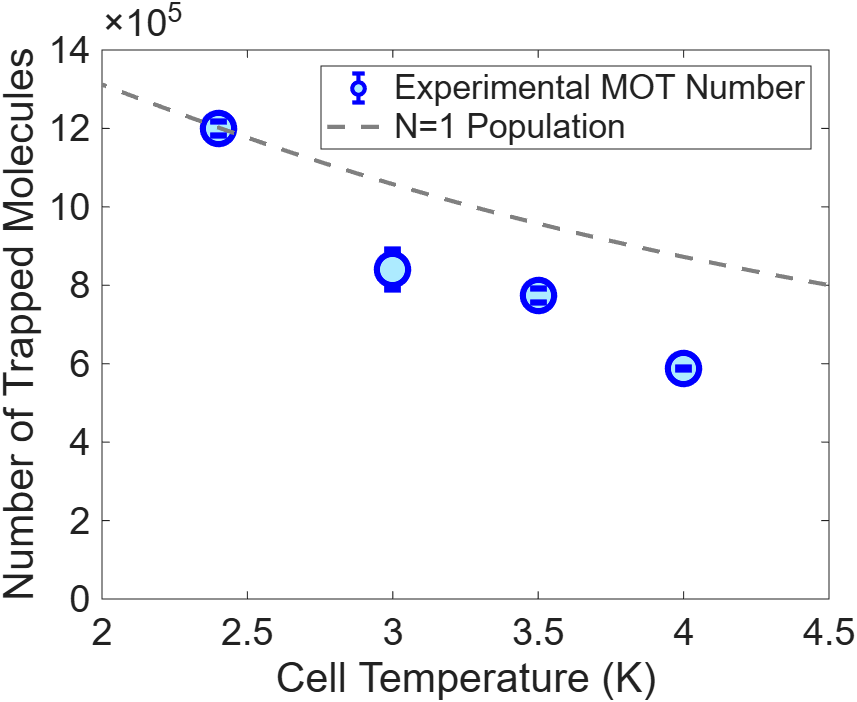}
    \caption{\textbf{Varying Cell Temperature.} Number of trapped molecules as a function of cell temperature for a two-stage cell with an 8-mm-diameter single-stage aperture, 10-mm-diameter second-stage aperture and 5.5\,\mbox{mm} spacing. This data was taken with a target holder length of -5.5\,\mbox{mm}. Error bars represent $\pm$1 standard error of the mean statistical uncertainties. The dashed line shows the theoretical N=1 population as a function of temperature, scaled to the 2.4\,\mbox{K} MOT brightness.}
    \label{fig:cell temps}
\end{figure}

We finally study the impact of the cell temperature on the number of molecules captured in the MOT. To increase the cell temperature, we fire a 450\,\mbox{nm}, 9\,\mbox{W} CW laser at the wall of the cell snorkel. The desired temperatures (2.4, 3, 3.5, and 4\,\mbox{K}) are achieved using pulse width modulation on the laser trigger, with 2.4\,\mbox{K} being the base temperature of the cell when this laser is off. We observe in Fig.~\ref{fig:cell temps} that the number of molecules captured decreases as the temperature increases. Two effects contribute to this decrease. First, the higher temperatures result in lower thermal populations of the N=1 state used for laser cooling. However, this decrease in N=1 population does not fully explain the reduction in the MOT number. The second effect is consistent with the increase in the forward velocity of the molecular beam at higher temperatures~\cite{White2024}.

\section{Conclusion}

We have optimized a cryogenic buffer gas beam source for loading a CaF MOT using the number of trapped molecules as the figure of merit. By correlating MOT loading with beam velocity measurements over many source configurations, we find that captured molecules leave the source with forward velocities between $\sim$110 and 140\,\mbox{m/s}.

We scanned various cell parameters for both single and two-stage sources to determine which produced the brightest MOT. For a single-stage cell, an 8\,\mbox{mm} aperture produced the largest MOT. Adding a second stage with an optimal gap of 5.5\,\mbox{mm} increased the MOT number by a factor of 1.5 relative to the best single-stage cell. We also observed that the position of the calcium target in the cell affected the number of molecules loaded into the MOT. Additionally, we found that the MOT number decreases by a factor of 2 as the cell temperature increases from 2.4 to 4\,\mbox{K}. The reduced thermal population of the N=1 level accounts for only part of this decrease. We attribute the remainder to the faster beam produced in a warmer cell, in line with the strong temperature dependence of the slow flux observed in previous studies~\cite{White2024}. Following these optimizations, we obtained a MOT with an average of 1.6 million molecules (80 - 98th percentile of 10 hours of running) and up to 1.9 million molecules in a 50-shot average. 

Our hybrid DSMC and Monte Carlo simulations reproduce the trends observed experimentally. These results point to several routes for further gains. First, the simulations predict that nearly all of the molecules reaching the MOT region are produced near the cell axis. Producing molecules in this region, by mounting the target on the back wall of the cell, for example, may further increase the MOT number. We note that these simulations do not include any dynamics present within the ablation process, the heated fill line, the three-dimensional flow around the target, or the time-dependent thermal and pressure dynamics within the cell. Experimentally, we find that the MOT number changes by more than a factor of 2 across the target positions tested (Fig.~\ref{fig:target position experiment}). This suggests that where the molecules are produced plays an important role and should be considered in the design of future sources. Second, due to the good agreement between simulation and experiment observed here, a broader optimization of the cell geometry could be carried out in simulation. The higher effective helium flow rate after ablation implied by the simulations indicates a flow regime where nozzles could be used to optimize MOT numbers~\cite{Xiao2019}. Third, helium reflected from the 3\,\mbox{K} shield strongly reduces the number of molecules captured in the MOT, so the source and shield geometry, such as the size and position of the shield aperture, should be optimized to reduce this effect. Finally, since the MOT loads only the slow tail of the beam, slowing the faster molecules would increase the trapped number. At fixed deceleration, the stopping distance scales as $v^2$, so simply extending the slowing distance is not beneficial due to the beam divergence. However, combining an increased slowing distance with transverse laser cooling~\cite{Nasir2026} may allow a larger fraction of the beam to be captured.

The optimizations and approach used here should apply to other buffer gas beam sources used for laser-coolable species, including polyatomic molecules.

\begin{acknowledgments}
We acknowledge experimental contributions to the apparatus by Jiahe Cao and Shi Zhang. We would like to thank Nick Hutzler for valuable discussions and careful reading of the manuscript. We are grateful to Curt Wittig and the Department of Chemistry at the University of Southern California for providing temporary laboratory space in which this research was conducted. Claude Opus 5.5 and ChatGPT 5 were used to debug and check for errors in the simulation code. All changes were reviewed and verified by the authors.
\end{acknowledgments}
\bibliographystyle{apsrev4-2}
\bibliography{CBGP_paper}

\clearpage

\appendix

\section{Correlation}
\label{supp:correlation}
\begin{figure}[htbp]
        \centering
        \includegraphics[width=0.9\linewidth]{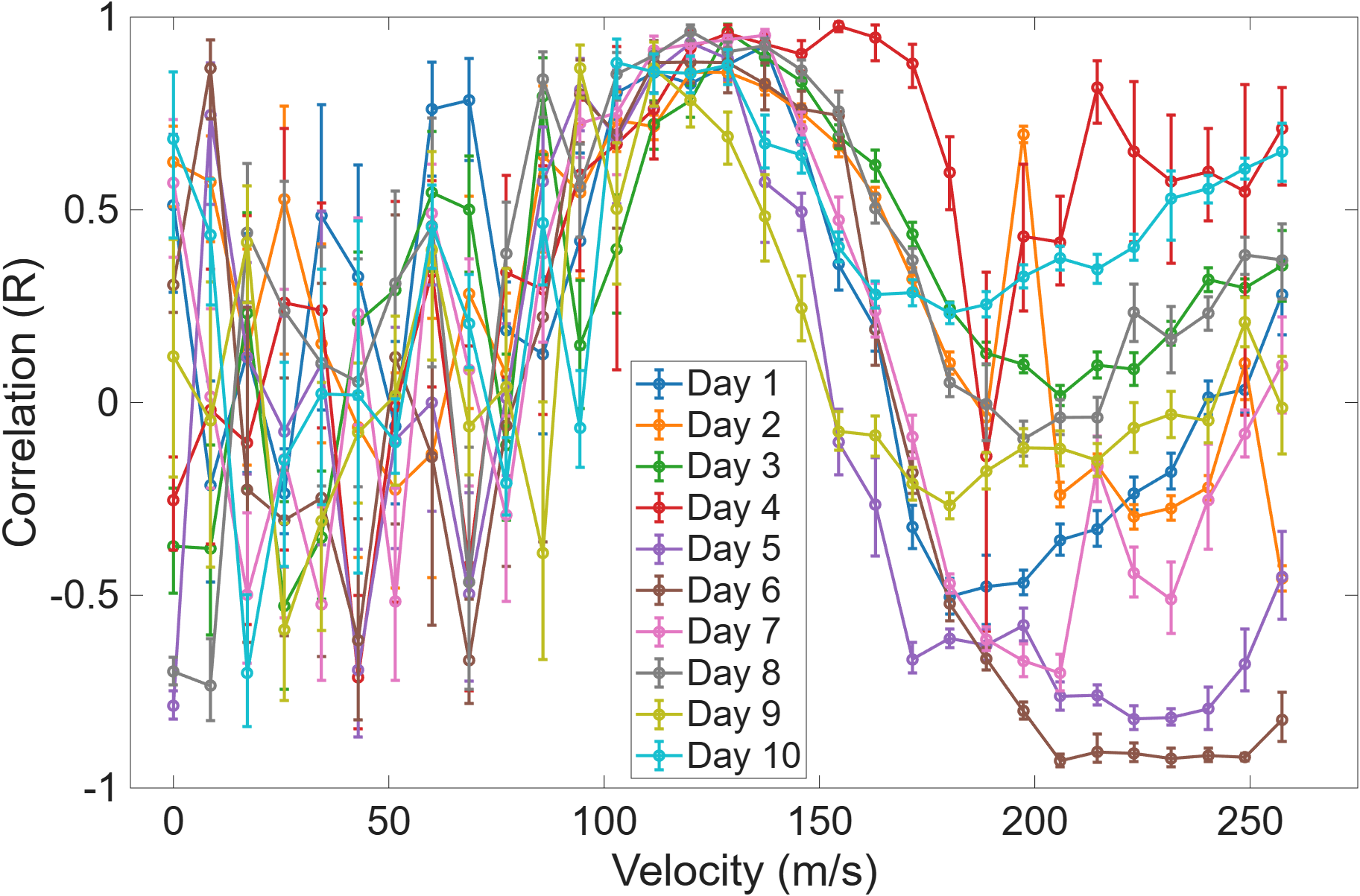} 
        \caption{\textbf{Individual Correlations.} Correlation between MOT number scans and forward velocity molecular flux scans plotted as a function of velocity for different cell configuration scanning days. Day 2 shows the correlations from the data set shown in Figures~\ref{fig:m vs flow example} and~\ref{fig:v_scan_example}.}
        \label{fig:all correlations vs v}
    \end{figure}

To calculate the correlation between the MOT number versus helium flow rate scans and molecular flux versus helium flow rate scans for different velocities, we calculate the Pearson coefficient,
\begin{equation}
\rho(M,V)=\frac{1}{N-1}\sum_{i=1}^{N}
\left(
\frac{M_i-\mu_M}{\sigma_M}
\right)
\left(
\frac{V_i-\mu_V}{\sigma_V}
\right)
\label{eq:correlation}
\end{equation}
where $M$ is the number of trapped molecules in the MOT, $V$ is the number of molecules imaged during the forward velocity molecular flux scan, and $\mu$ and $\sigma$ are the corresponding mean and standard deviation of the two data sets. For each pair of scans, we also found the upper and lower confidence bounds using a 68\% confidence interval. These correlations and their confidence intervals are plotted in Fig.~\ref{fig:all correlations vs v}. Fig.~\ref{fig:v vs correlation} is then generated by performing Fisher transformations on the correlations and their confidence bounds and then averaging them. New confidence bounds are generated, also with a 68\% confidence interval.

\section{Simulation}
\label{supp:simulation}
\begin{figure}[htbp]
        \centering
        \includegraphics[width=0.8\linewidth]{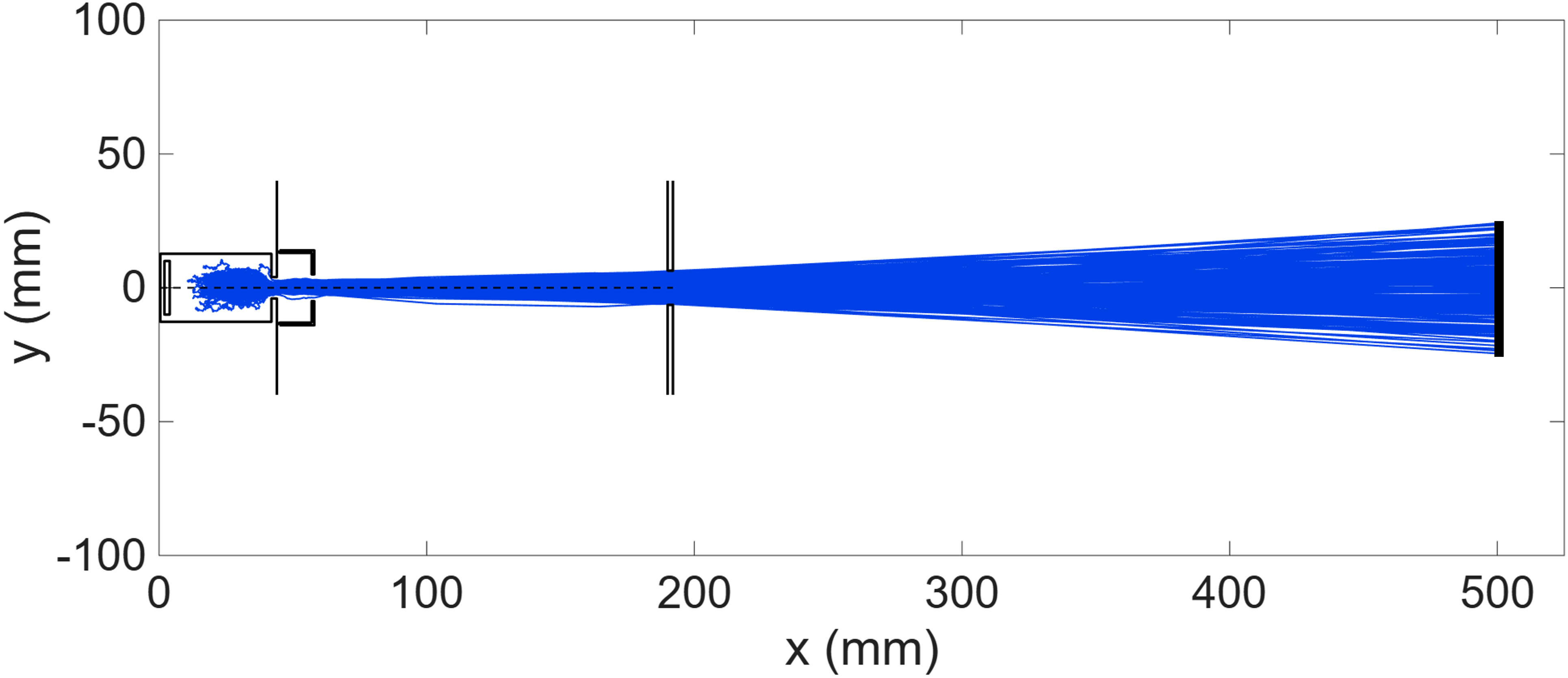} 
        \caption{\textbf{Simulation Geometry.} Geometry of a two-stage cell DSMC simulation with traces plotted for molecules with velocities 108.5 - 137.5\,\mbox{m/s} that reach the detector. The first vertical line indicates the single-stage aperture. The second shows the 3\,\mbox{K} shield and its aperture. The final vertical line represents the detector. Molecules that pass through the single-stage aperture are considered ``extracted" and molecules that pass through all apertures and hit the detector are considered to have ``reached the MOT region."}
        \label{fig:mot detector}
\end{figure}

The simulation geometry is shown in Fig.~\ref{fig:mot detector}. In this paper, a detector radius of 2.5\,\mbox{cm} is used. While this is larger than the MOT beams used in this experiment by about a factor of 2, we saw no appreciable difference in running with this region as opposed to a 1.1\,\mbox{cm} radius region. We opted to use this larger region to keep the simulation times manageable. For most simulations in the paper, $\sim$50,000 CaF molecules are launched on axis, $\sim$20\,\mbox{mm} from the single-stage cell aperture, with a 3\,\mbox{mm} Gaussian width. 

\vspace{4mm}

\begin{figure}[htbp]
    \begin{subfigure}[t]{0.48\columnwidth}
        \centering
        \includegraphics[width=\linewidth]{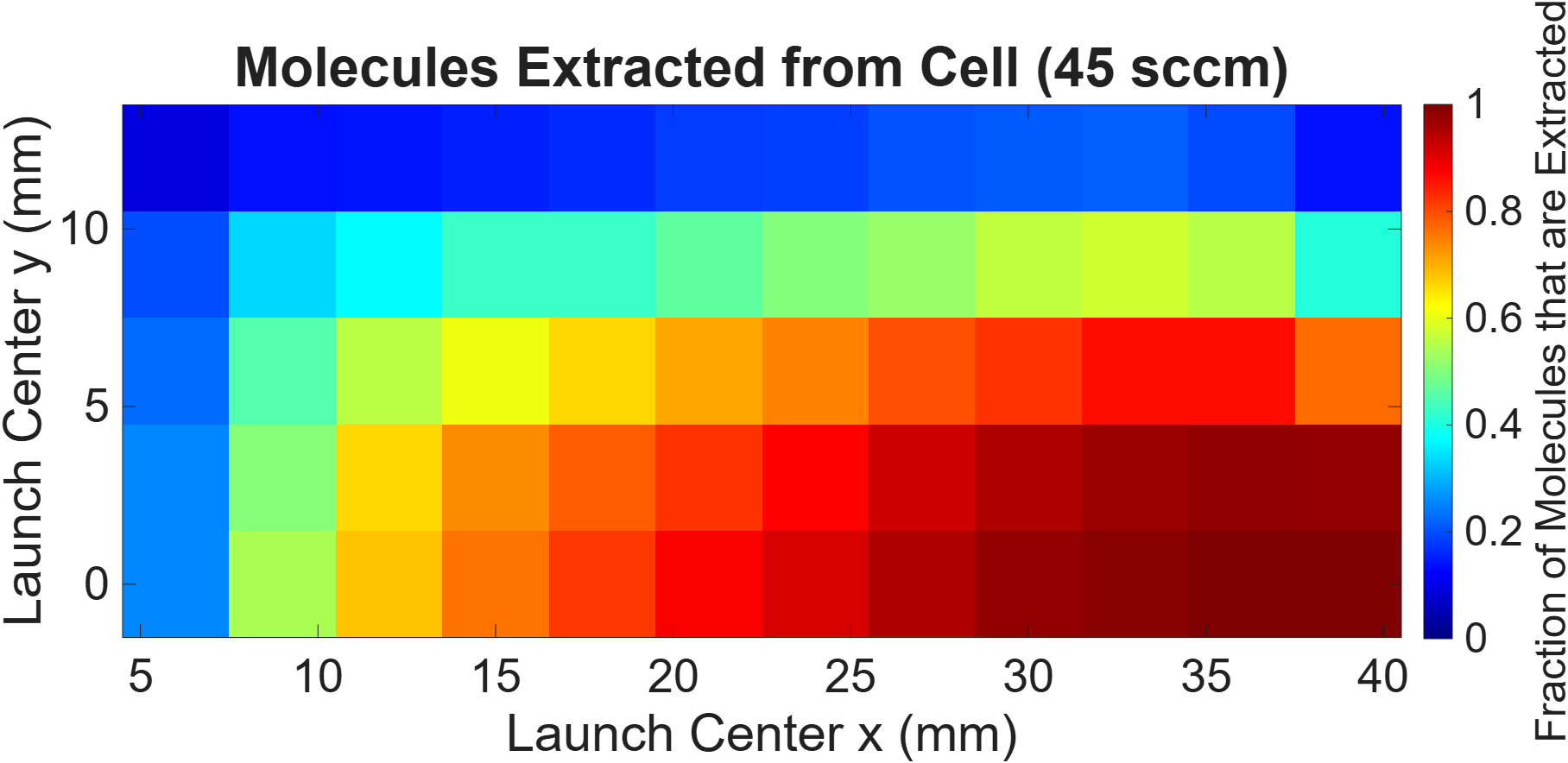}
         \caption{}
        \label{fig:grid compare extract 1.5}
    \end{subfigure}
    \hfill
    \begin{subfigure}[t]{0.5\columnwidth}
        \centering
        \includegraphics[width=\linewidth]{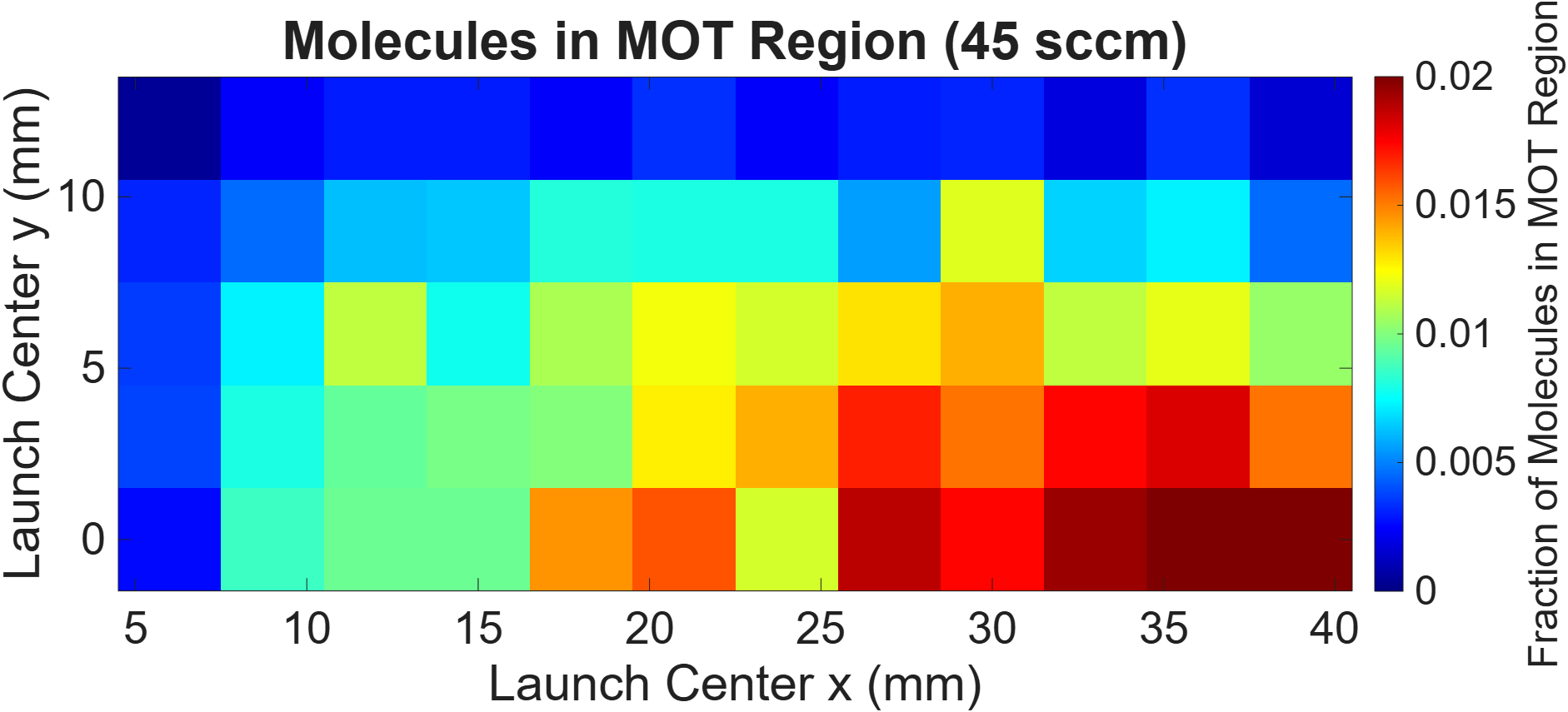}
               \caption{}
        \label{fig:grid compare detect 1.5}
    \end{subfigure}

    \medskip
    
    \begin{subfigure}[t]{0.48\columnwidth}
        \centering
        \includegraphics[width=\linewidth]{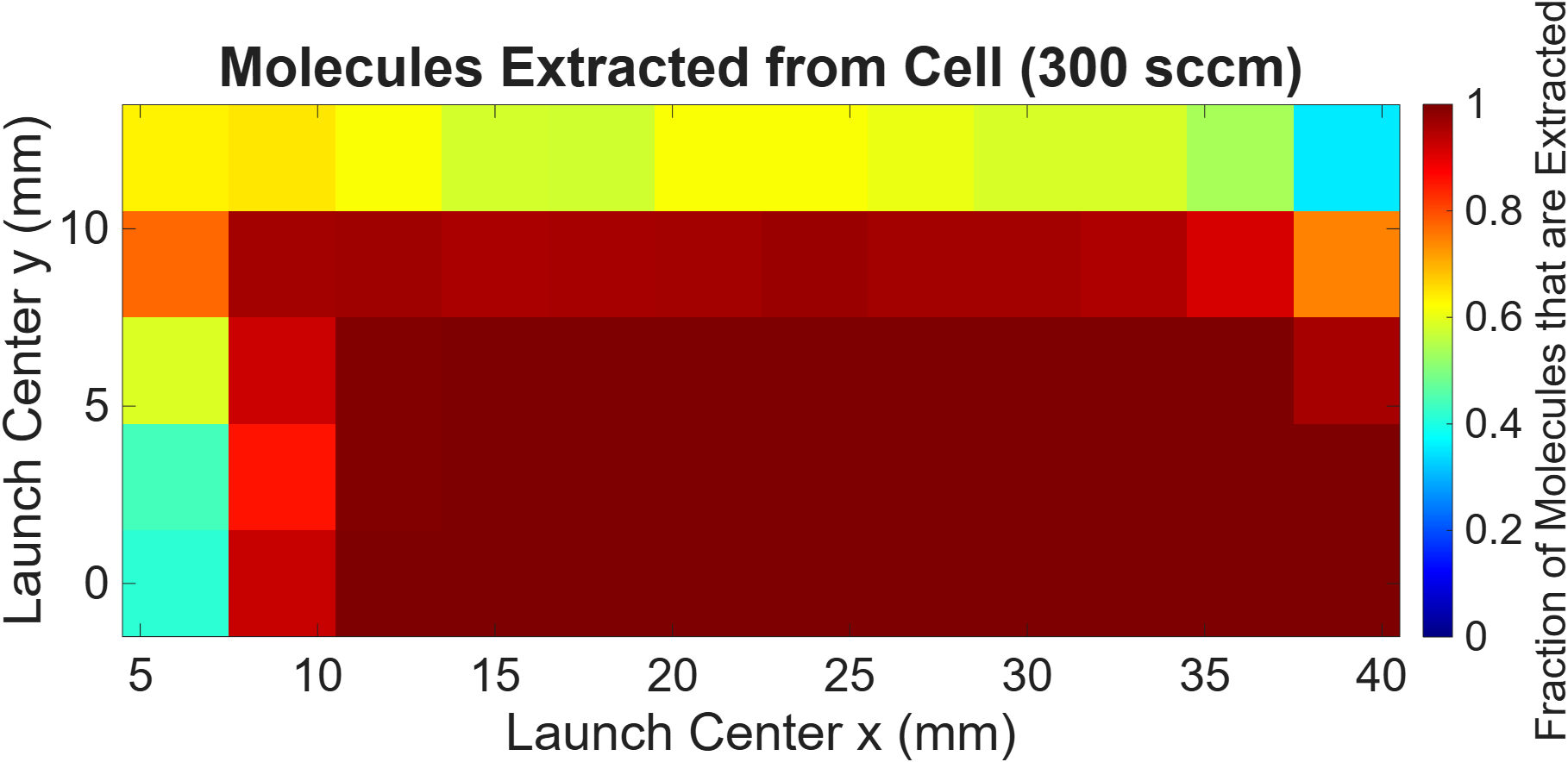}
         \caption{}
        \label{fig:grid compare extract 10}
    \end{subfigure}
    \hfill
    \begin{subfigure}[t]{0.5\columnwidth}
        \centering
        \includegraphics[width=\linewidth]{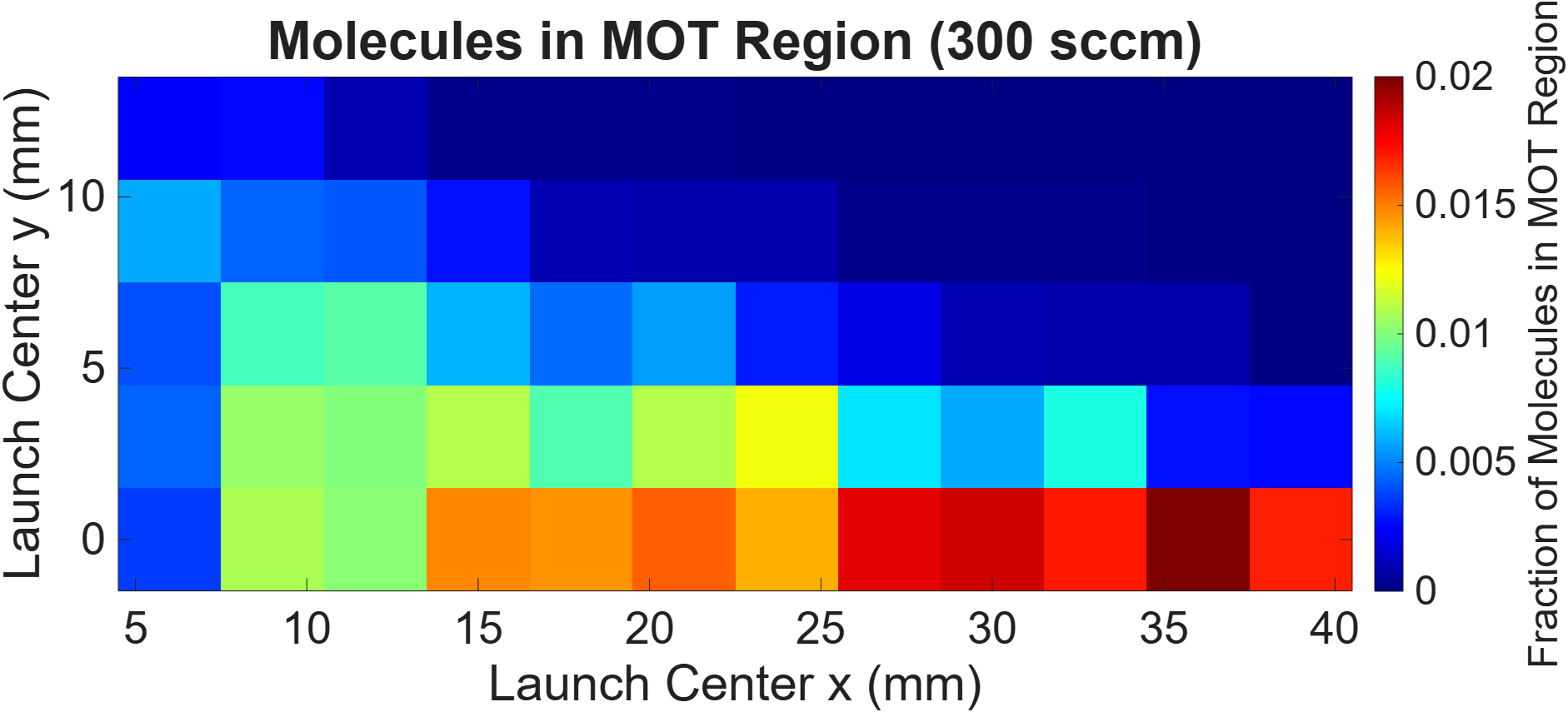}
               \caption{}
        \label{fig:grid compare detect 10}
    \end{subfigure}
    \caption{\textbf{Changing the Helium Source Location.} (a) Fraction of launched CaF molecules that exit the single-stage cell aperture as a function of their initial positions. Here, we use a single-stage cell geometry with an 8\,\mbox{mm} aperture and a 45\,\mbox{sccm} effective (equivalent to 1.5\,\mbox{sccm} continuous) helium flow rate. Helium is flowed in uniformly from all cell walls, excluding the front aperture. (b) Fraction of launched CaF molecules that reach the MOT region as a function of their initial positions. Here, we use a single-stage cell geometry with an 8\,\mbox{mm} aperture and a 45\,\mbox{sccm} effective (equivalent to 1.5\,\mbox{sccm} continuous) helium flow rate. Helium is flowed in uniformly from all cell walls, excluding the front aperture. (c), (d) Same as (a) and (b) but using a 300\,\mbox{sccm} effective (equivalent to 10\,\mbox{sccm} continuous) helium flow rate.}
    \label{fig:grid compare}
    \end{figure}

As discussed in Sec.~\ref{simulations}, the helium may be predominantly introduced into the cell from the cell walls. To characterize this effect, we reran the grid scans with the helium introduced from the cell walls (Fig.~\ref{fig:grid compare}). No significant difference is observed when compared with Fig.~\ref{fig:caf-extraction}. 

\begin{figure}[htbp]
        \centering
        \includegraphics[width=0.8\linewidth]{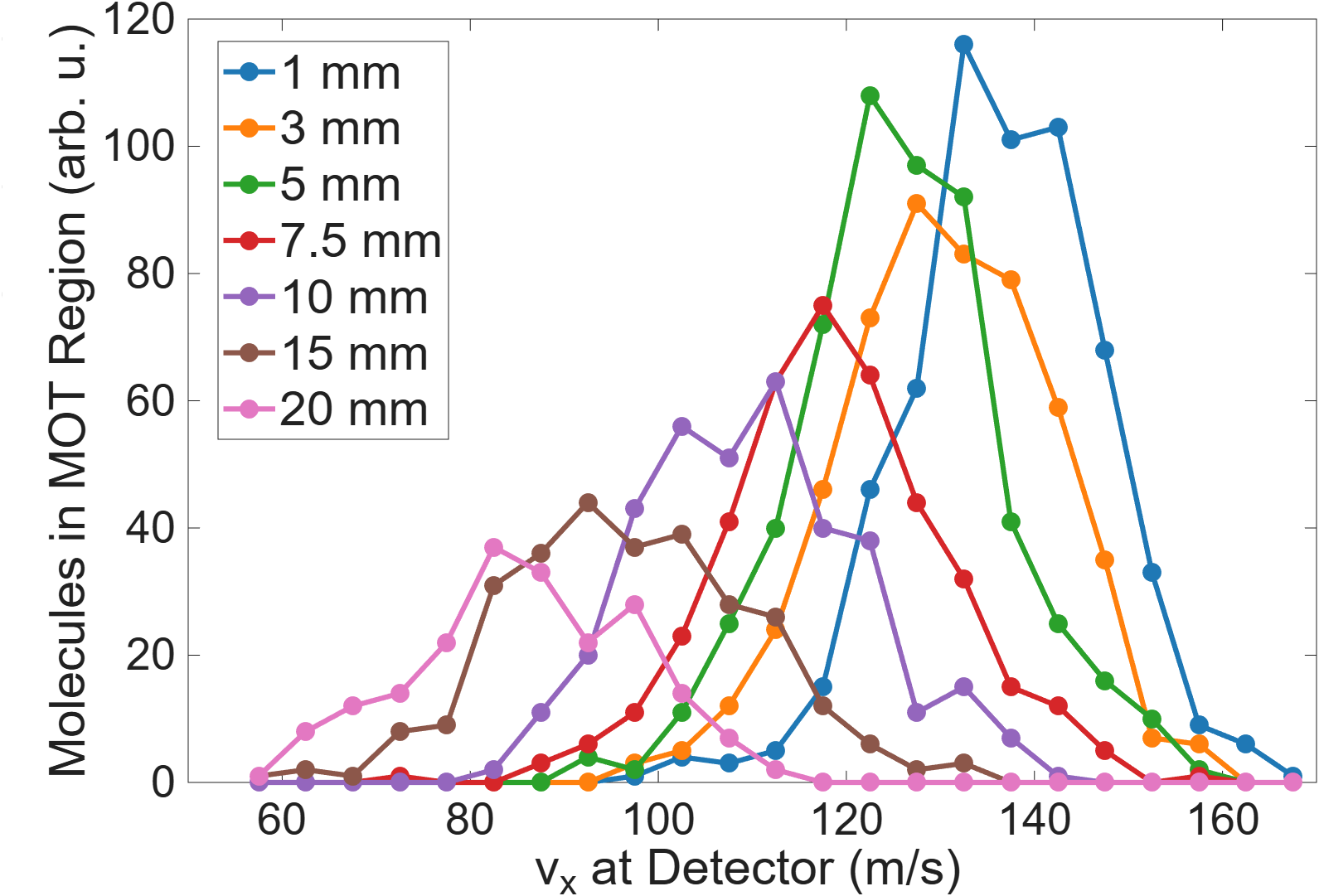} 
        \caption{\textbf{Two-Stage Simulation for All Velocities.} DSMC simulation of molecules that enter the MOT region with an effective helium flow rate of 120\,\mbox{sccm} for 1, 3, 5, 7.5, 10, 15, and 20\,\mbox{mm} two-stage cell spacings plotted against the final forward velocity of those molecules.}
        \label{fig:sim results all velocities}
\end{figure}

For each cell configuration simulation scan, we measure the number of molecules that reach the MOT region and note their final forward velocities. Fig.~\ref{fig:sim results all velocities} shows the results for the two-stage spacing scan with an effective helium flow rate of 120\,\mbox{sccm}. To generate Fig.~\ref{fig:2 stage spacing simulation}, we plot only the molecules with forward velocities from 108.5 - 137.5\,\mbox{m/s}. We use the same procedure for Figures~\ref{fig:mot scan simulation} and~\ref{fig:1 stage aperture simulation}.

\section{Data Thresholding}
\label{supp:data selection}

\begin{figure}[htbp]
    \begin{subfigure}{0.23\textwidth}
        \centering
        \includegraphics[width=\linewidth]{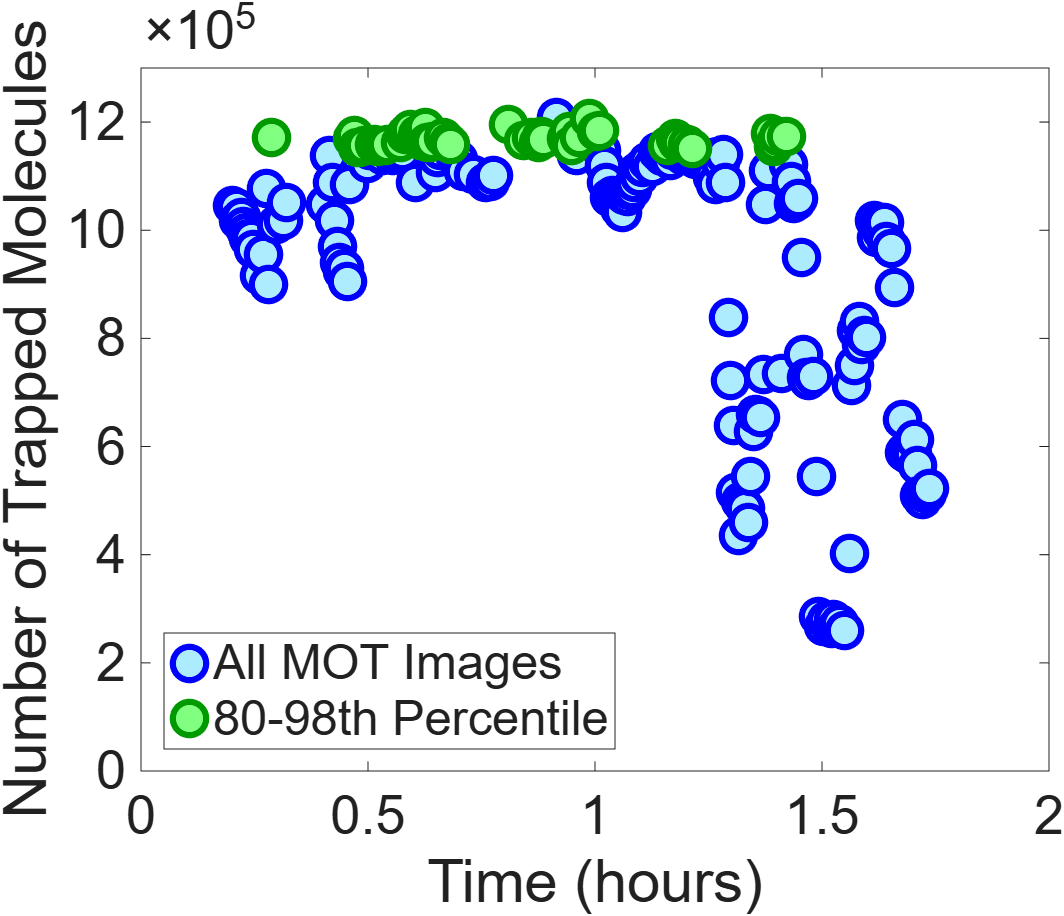}
         \caption{}
        \label{fig:daily run 1}
    \end{subfigure}
    \begin{subfigure}{0.23\textwidth}
        \centering
        \includegraphics[width=\linewidth]{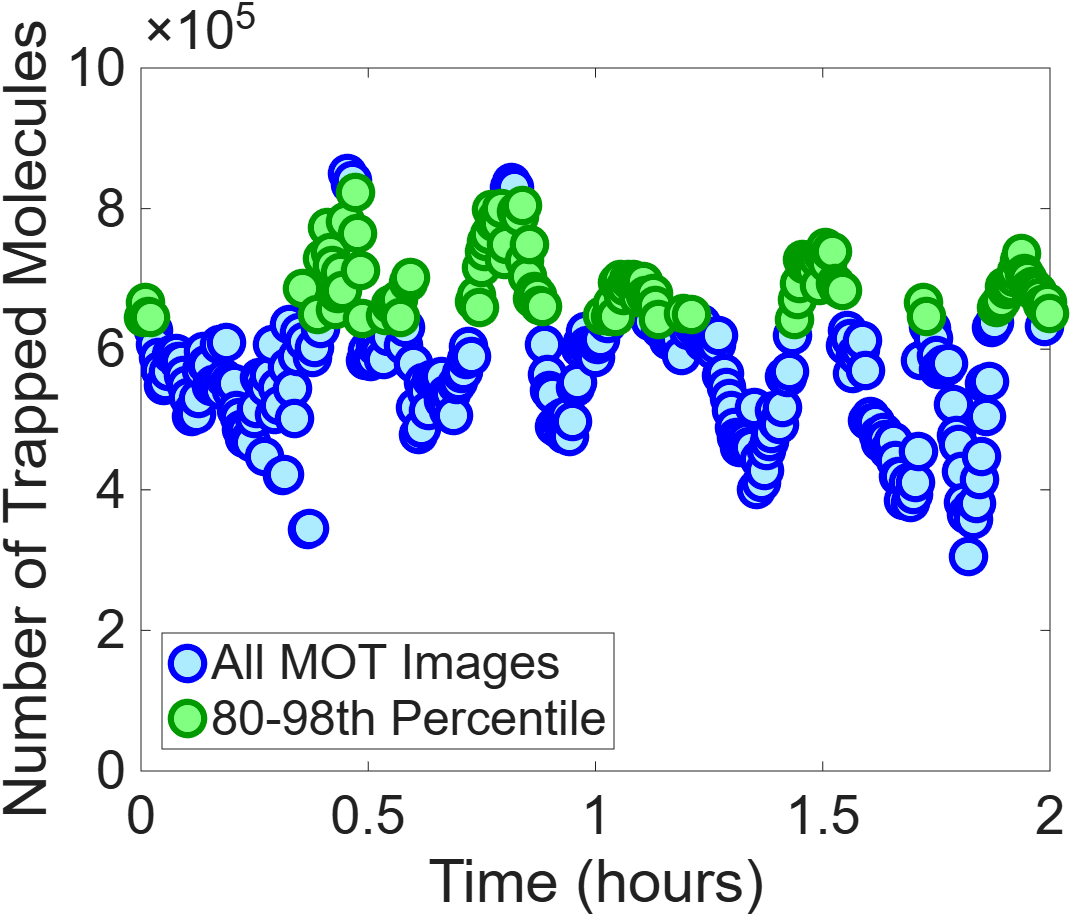}
               \caption{}
        \label{fig:daily run 2}
    \end{subfigure}
    \caption{\textbf{MOT Number Thresholding.} Number of molecules trapped in the MOT plotted as a function of time. Green points show the images that are included in the 80 - 98th percentile range and averaged to provide the final MOT number. Blue points show all other images not within the selected percentile range. (a) Section of a data set which is fairly stable but includes some parameter scans. This data set was monitored, and frequencies and couplings were checked routinely. (b) Section of a data set which oscillates over time. This data set was not monitored, and oscillations can be attributed to frequency and coupling drifts.}
    \end{figure}

For each configuration, we average the MOT number over the 80 - 98th percentile of the images in that data set, as shown in Figures~\ref{fig:daily run 1} and~\ref{fig:daily run 2}. Green points show the images included in this average, and blue points show all other images. Fig.~\ref{fig:daily run 1} shows a data set which is fairly stable for a period of time. This data set was being monitored while imaging (frequencies adjusted, laser powers monitored, etc.). Fig.~\ref{fig:daily run 2} shows a data set with oscillations. This data set was not monitored, and we attribute the oscillation to drifts in our X-A laser frequency over time, as well as changes in coupling due to room temperature shifts. Averaging over all shots would mix these technical drifts into the comparison between configurations. Alternatively, quoting only the maximum of each data set would be susceptible to noise and have poor statistics. The range used better represents the MOT number we obtain in daily operation.

\end{document}